\documentclass[%
 amsmath,amssymb,
 aps,
 prfluids,
 onecolumn,
]{revtex4-2}

\usepackage{graphicx}
\usepackage{dcolumn}
\usepackage{bm}
\usepackage{hyperref}

\graphicspath{{figures/}} 

\begin{document}

\preprint{APS/123-QED}

\title{Modelling bubble collapse anisotropy in complex geometries}

\author{Elijah D. Andrews}
 \email{e.d.andrews@soton.ac.uk}
\author{Ivo R. Peters}%
\affiliation{%
 Faculty of Engineering and Physical Sciences, University of Southampton, Southampton, SO17 1BJ, UK
}%

\date{\today}

\begin{abstract}
A gas or vapor bubble collapsing in the vicinity of a rigid boundary displaces towards the boundary and produces a high-speed jet directed at the boundary. This behavior has been shown to be a function of the `anisotropy' of the collapse, measured by a dimensionless representation of the Kelvin impulse known as the anisotropy parameter [Supponen \emph{et al.}, J. Fluid Mech. 802, 263-293 (2016)]. However, characterisation of the anisotropy parameter in different geometries has been limited to simplified analytic solutions. In this work we develop an inexpensive numerical model, based on the Boundary Element Method, capable of predicting the anisotropy parameter for any rigid complex geometry. We experimentally explore a robust measure of bubble displacement, showing that the bubble displacement in a range of complex geometries behaves as a single function of the predicted anisotropy parameter values.
\end{abstract}

\maketitle


\section{Introduction}
Collapsing bubbles have been studied in great depth for many years due to their ubiquity in both nature and engineering. Common applications include cavitation damage \citep{Rayleigh1917, Luo2016, Lu2016, Sagar2020}, cleaning \citep{Ohl2006, Verhaagen2016, Birkin2016, Chahine2016, Verhaagen2016a, Reuter2017}, and various biomedical applications \citep{DeLaTorre1992, Palanker2002, Brennen2003, Canchi2017, OyarteGalvez2020}. Although many such applications feature numerous bubbles collapsing, it is important to study the fundamental behavior of individual bubbles to identify how key parameters affect the collapse behaviour. This study goes as far back as \citet{Rayleigh1917} where a single spherical void collapsing in an infinite fluid was treated. The study of single bubbles collapsing has since grown significantly, addressing a wide range of questions. Studies have experimentally and numerically investigated a wide variety of bubble collapse properties such as bubble collapse shapes \citep{Blake1986, Brujan2019, Zhang2020}, induced pressures \citep{Benjamin1966, Li2015, Li2016, Supponen2019}, boundary shear stresses \citep{Dijkink2008a, Koukouvinis2018, Zeng2018, Gonzalez-Avila2020, Zeng2022}, and shock wave properties \citep{Tagawa2016, Sinibaldi2019, Trummler2021}. Much of this research focuses on simple boundary geometries such as flat plates, free surfaces, constant pressure gradients, and other boundaries for which analytic solutions of the flow field exist. In recent years, there has been increasing interest in complex geometries. For example, studies have explored jet direction for various complex geometries \citep{PhysRevFluids.3.081601, Molefe2019, Andrews2020}, bubble dynamics near curved rigid surfaces \citep{TOMITA2002}, dynamics in combinations of boundaries and free surfaces \citep{Zhang2017, Quah2018, Kiyama2021}, jetting and shear stress between two walls \citep{Han2018, Gonzalez-Avila2020}, and bubble shape variation at the corner of a rigid wall \cite{Zhang2020}.

Many parameters of the dynamics of a bubble collapsing near a flat plate depend on the standoff distance $\gamma = Y / R_0$ where $Y$ is the distance from a boundary or fluid interface and $R_0$ is the maximum bubble size \citep{Kroninger2010, Supponen2016, Zeng2022}. However, for more complex geometries, it is difficult to define a geometric equivalent to the standoff parameter. Even with very limited complexity, such as in a corner, it is not trivial to define such a parameter.

The Kelvin impulse is a parameter that captures the overall motion of the fluid. The Kelvin impulse is the net force acting on the fluid integrated over time, effectively the total fluid momentum. This parameter was described well by \citet{Benjamin1966} and has been studied analytically and numerically for many years \citep{Blake1982, Blake1986, Kucera1990, Harris1996, Blake2015}. More recently, \citet{Supponen2016} presented a dimensionless version of the Kelvin impulse called the anisotropy parameter, denoted $\zeta$. Analytic solutions for the anisotropy parameter were derived for a number of different sources of anisotropy and many bubble collapse properties have been shown to be functions of the anisotropy parameter, regardless of the source of anisotropy. For example, jet speed, bubble displacement, jet volume \citep{Supponen2016}, shockwave energy \citep{Supponen2017}, and rebound energy \citep{Supponen2018}. Thus, the anisotropy parameter is a powerful tool for predicting bubble collapse behaviour.

\citet{Harris1996} presented a simple boundary element model for estimating the Kelvin impulse for a bubble near a complex geometry. In the current research, we present a similar model that predicts the Kelvin impulse for complex geometries. We extend this model by non-dimensionalizing the results to produce the anisotropy parameter following the example of \citet{Supponen2016}. With this model, anisotropy parameter values can be predicted for complex geometries whereas they were previously only studied for geometries with limited analytic solutions. The study of bubble collapse properties as functions of anisotropy magnitude can thus be extended to complex geometries.

\section{Methods}
\label{sec:methods}
In this section, the procedure for calculating the general Kelvin impulse is derived and the procedure for converting this to the anisotropy parameter is presented. These calculations rely on numerical solutions to the Rayleigh-Plesset equation so various formulations of the Rayleigh-Plesset equation are compared. Three methods for computing the anisotropy parameter for a range of geometries are then presented. Finally, the data analysis techniques used to process experimental results are outlined.
\subsection{Computing the anisotropy parameter}
The Kelvin impulse, $\mathbf{I}(\tau)$, of a fluid during the time period $0 < t < \tau$ can be represented as the integral of the force acting on the fluid over time,
\begin{equation}
\mathbf{I}(\tau) = \int_0^\tau \mathbf{F}(t) dt,
\label{eq:impulse_integral_equation}
\end{equation}
where $\tau$ is an arbitrary point in time and the force $\mathbf{F}(t)$ acting on the fluid is equal and opposite to the force the fluid exerts on the bounds of the domain. For a bubble in an infinite fluid bounded only by a rigid geometry, and following the derivation of \citet{Blake1982}, this force can be written as
\begin{equation}
\mathbf{F}(t) = -\rho \int_S \left[ \frac{1}{2} |\nabla\phi|^2 \mathbf{n} - \frac{\partial\phi}{\partial n} \nabla\phi \right] dS,
\label{eq:original_kelvin_impulse_force}
\end{equation}
where $\rho$ is the surrounding fluid density, $S$ is the boundary surface, and $\mathbf{n}$ is a vector normal to the surface $S$. Potential flow is assumed with velocity potential $\phi$ and $\partial\phi / \partial n$ is the derivative of the potential in the normal direction. Note here that the sign varies from previous works \citep{Blake1982, Blake1988, Harris1996} where the normal vector is defined as outward from the domain. In Eq. (\ref{eq:original_kelvin_impulse_force}) the normal vector is defined as inward to the domain in accordance with the convention used in this work for the boundary element method.

On the surface of the boundary, the normal velocity $\partial \phi / \partial n = 0$ \citep{Harris1996} and thus the force reduces to the integral of the dynamic pressure on the boundary
\begin{equation}
\mathbf{F}(t) = -\frac{1}{2}\rho \int_S |\nabla\phi|^2 \mathbf{n} dS.
\label{eq:kelvin_impulse_force}
\end{equation}

The bubble and all boundaries are modelled using point sinks. The strength of these sinks scales directly with the bubble sink strength $m_b(t)$ which varies with time. Thus, we can express the velocity $\nabla\phi$ induced at a position $j$ by a boundary element sink as the bubble sink strength multiplied by the equivalent velocity $\nabla\phi'$ when the bubble sink strength is $m_b = 1$ m$^3$/s. Properties computed with $m_b = 1$ m$^3$/s are denoted by a prime. As such this velocity is denoted $\nabla\phi'$ and we can write 
\begin{equation}
\nabla\phi|_j = m_b(t) \nabla \phi'|_j.
\end{equation}

We therefore express the force as
\begin{equation}
\mathbf{F}(t) = -\frac{1}{2}\rho \int_S |m_b(t)\nabla\phi_j'|^2 \mathbf{n} dS = m_b(t)^2 \left[-\frac{1}{2}\rho \int_S |\nabla\phi_j'|^2 \mathbf{n} dS\right] = m_b(t)^2 \mathbf{F}',
\label{eq:force_vs_force_prime_relation}
\end{equation}
such that
\begin{equation}
\mathbf{F}' = -\frac{1}{2}\rho \int_S |\nabla\phi_j'|^2 \mathbf{n} dS,
\label{eq:F_prime_definition}
\end{equation}
where $\mathbf{F}'$ is the equivalent force for a bubble sink strength of $m_b = 1$ m$^3$/s. $\mathbf{F}'$ is time-independent, purely depending on the geometry and bubble position, which is assumed to be fixed. Different methods for calculating $\mathbf{F}'$ are presented in Secs. \ref{sec:analytic_kelvin_impulse} to \ref{sec:bem_kelvin_impulse}.

Thus, combining Eqs. (\ref{eq:impulse_integral_equation}) and (\ref{eq:force_vs_force_prime_relation}), the impulse integral equation becomes
\begin{equation}
\mathbf{I}(\tau) = \mathbf{F}' \int_0^\tau m_b(t)^2 dt.
\end{equation}

By assuming that the bubble remains spherical throughout the collapse, and defining the bubble sink strength to be the rate of change of bubble volume $V$, the sink strength is given by
\begin{equation}
m_b(t) = \frac{dV}{dt} = 4\pi R^2 \dot{R}
\end{equation}
where $R$ is the bubble radius and $\dot{R}$ is the time-derivative of bubble radius.

The Kelvin impulse therefore becomes
\begin{equation}
\mathbf{I}(\tau) = \mathbf{F}'\int_0^\tau \left[4 \pi R^2 \dot{R} \right]^2 dt = 16 \pi^2 \mathbf{F}' \int_0^\tau R^4 \dot{R}^2 dt.
\label{eq:numerical_kelvin_impulse}
\end{equation}
which can be computed using a numerical solution to the Rayleigh-Plesset equation.

A scaling between the anisotropy parameter $\boldsymbol{\zeta}$ and Kelvin impulse $\mathbf{I}$ was presented by \citet{Supponen2016}, 
\begin{equation}
\mathbf{I} = 4.789 R_0^3 \sqrt{\Delta p \rho} \boldsymbol{\zeta}
\label{eq:anisotropy_scaling}
\end{equation}
where $R_0$ is the initial radius of the bubble, which is taken to be its maximum size; $\Delta p$ is the difference between the pressure at an infinite distance and the internal pressure of the bubble; and $\rho$ is the density of the liquid. This relation was analytically derived from the initial expansion and subsequent collapse of a bubble collapsing in a pressure gradient.

Thus, Eqs. (\ref{eq:numerical_kelvin_impulse}) and (\ref{eq:anisotropy_scaling}) can be combined to determine the anisotropy parameter for any geometry,
\begin{equation}
\boldsymbol{\zeta} = \frac{16 \pi^2 \mathbf{F}'}{4.789 R_0^3 \sqrt{\Delta p \rho}} \int_0^\tau R^4 \dot{R}^2 dt
\label{eq:anisotropy}
\end{equation}
where $\tau$ is now the duration of the initial expansion and collapse.

The vector anisotropy parameter $\boldsymbol{\zeta}$ represents both a magnitude and direction. The direction is equivalent to that previously used to study jet direction in complex geometries \citep{PhysRevFluids.3.081601, Molefe2019, Andrews2020}. In this work we focus on the magnitude, represented by the scalar anisotropy parameter $\zeta$.

\subsection{Formulations of the Rayleigh-Plesset equation}
In order to numerically solve the integral in Eq. (\ref{eq:anisotropy}), the radius $R$ and radial velocity $\dot{R}$ must be known. In order to compute these we use the Rayleigh-Plesset equation which has long been the standard model of bubble collapse. Various modifications have been derived for differing conditions and assumptions. By assuming that the bubble collapse is primarily inertial, in an incompressible liquid, with no heat transferred across the bubble boundary, and with internal gas that behaves isentropically \citep{Brennen1995}, the Rayleigh-Plesset equation is
\begin{equation}
\frac{-\Delta p}{\rho_L} + \frac{p_{G_0}}{\rho_L} \left(\frac{R_0}{R}\right)^{3k} = R\ddot{R} + \frac{3}{2}\dot{R}^2 + \frac{4 \nu_L}{R}\dot{R} + \frac{2s}{\rho_L R}
\label{eq:typical_rp_formulations}
\end{equation}
where $\Delta p$ is the driving pressure of the collapse, often defined as $p_\infty - p_V$ where $p_V$ is the vapor pressure of water and $p_\infty$ is the far-field pressure; $p_{G_0}$ is the initial pressure of gas inside the bubble; $\rho_L$ and $\nu_L$ are the density and kinematic viscosity of the liquid, respectively; $s$ is the surface tension; and $k$ is the ratio of specific heats of water vapor $k = c_p / c_v \approx 1.33$.

A derivation of the Rayleigh-Plesset equation was presented by \citet{Harris1996} that includes the effect of a nearby rigid wall and the buoyancy of the bubble,
\begin{equation}
\frac{- p_\infty}{\rho_L} + \frac{p_{G_0}}{\rho_L} \left(\frac{R_0}{R}\right)^{3k} + gz = R\ddot{R} + \frac{3}{2}\dot{R}^2 - \frac{\partial \phi_w}{\partial t}
\label{eq:harris_rp}
\end{equation}
where $\phi_w$ is the velocity potential induced by the wall at the bubble position; $g$ is the acceleration due to gravity; and $z$ is the depth of the bubble in the water.

Here we shall present a comparison between four different models and experimental data. The first model is Eq. (\ref{eq:typical_rp_formulations}), denoted `\ref{eq:rp-complete}'. The second model is the same, but neglecting the effects of viscosity and surface tension, denoted `\ref{eq:rp-inertial}'. The third model, denoted `\ref{eq:rp-wall-model}', is based on Eq. (\ref{eq:harris_rp}), however the effect of buoyancy  is neglected and $p_\infty$ is replaced by $\Delta p$ for consistent comparison with the other models. The final model, denoted `\ref{eq:rp-no-internal-gas}', assumes a constant pressure difference between the inside and outside of the bubble, effectively assuming that there is no internal gas to be compressed. This model is used by \citet{Obreschkow2012} in order to find analytical approximations to the solution. The equations for each of these models are as follows:

\begin{align}
\frac{-\Delta p}{\rho_L} + \frac{p_{G_0}}{\rho_L} \left(\frac{R_0}{R}\right)^{3k} &= R\ddot{R} + \frac{3}{2}\dot{R}^2 + \frac{4 \nu_L}{R}\dot{R} + \frac{2s}{\rho_L R} \tag{Complete} \label{eq:rp-complete} \\
\frac{-\Delta p}{\rho_L} + \frac{p_{G_0}}{\rho_L} \left(\frac{R_0}{R}\right)^{3k} &= R\ddot{R} + \frac{3}{2}\dot{R}^2 \tag{Inertial} \label{eq:rp-inertial} \\
\frac{-\Delta p}{\rho_L} + \frac{p_{G_0}}{\rho_L} \left(\frac{R_0}{R}\right)^{3k} &= R\ddot{R} + \frac{3}{2}\dot{R}^2 - \frac{\partial \phi_w}{\partial t} \tag{Wall model} \label{eq:rp-wall-model} \\
\frac{-\Delta p}{\rho_L} &= R\ddot{R} + \frac{3}{2}\dot{R}^2 \tag{No internal gas} \label{eq:rp-no-internal-gas}
\end{align}

\begin{figure}
\centerline{\includegraphics{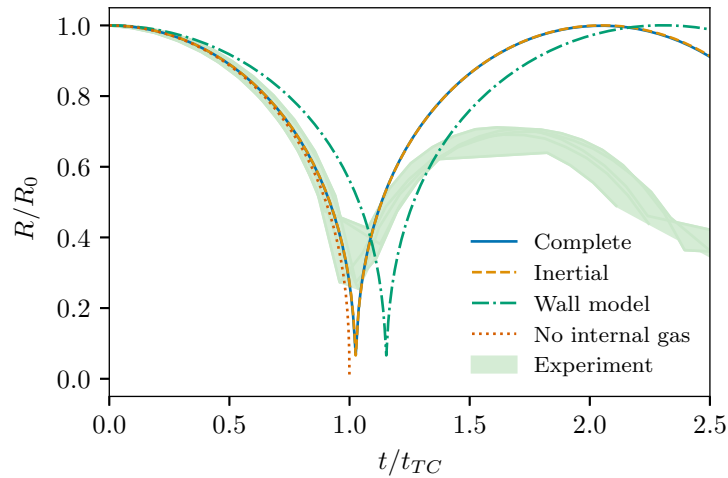}}
\caption{A comparison between the four Rayleigh-Plesset models and experimental data. The experimental data used here are a series of five bubbles generated in the same location near a slot, the green shaded area shows the spread of the data. The experimental mean of the radius and position at the maximum bubble size are used as the initial conditions for numerical models. Bubbles ranged in size from $R_0 = 1.13$ mm to $R_0 = 1.34$ mm. Radius is normalized with the initial radius $R_0$ and time is normalized with the Rayleigh collapse time $t_{TC}$ \citep{Rayleigh1917}.}
\label{fig:rp_models_comparison}
\end{figure}

Figure \ref{fig:rp_models_comparison} shows the comparison between these models and experimental data. For simple comparison, the models are initiated at maximum bubble size with internal pressure equal to the vapor pressure of water. For most models, and experimental data, the bubbles collapse to a minimum size and then rebound. Notably, around the minimum size the bubble collapses and rebounds very fast. Due to limitations in the temporal resolution of the high-speed camera, the experimental data do not accurately capture the minimum size of the bubble.

After rebound, all of the Rayleigh-Plesset models with internal gas recover very close to their initial size, whereas the experimental data shows that the bubbles do not achieve their original size due to significant energy being dissipated during the collapse.

Comparing the two standard models, one with viscosity and surface tension (`\ref{eq:rp-complete}') and one without (`\ref{eq:rp-inertial}'), it is clear that viscosity and surface tension are negligible for this regime of bubble collapse.

The model that includes the effect of a nearby wall (`\ref{eq:rp-wall-model}') consistently overestimates the collapse duration when compared to experimental data. This is likely due to the assumption of spherical symmetry in the model. In reality, the side of a bubble opposite to a nearby boundary does not get impeded by the boundary. However, the wall model assumes spherical symmetry, which means all sides of the bubble are considered to be equally impeded, resulting in a generally slower collapse.

In comparison with experimental data, the `\ref{eq:rp-complete}', `\ref{eq:rp-inertial}', and `\ref{eq:rp-no-internal-gas}' models all match well. In general, a benefit of the Kelvin impulse model presented in this work is that any model for the bubble radius variation with time can be used without significant methodological modifications. From here onwards, we utilize the `\ref{eq:rp-inertial}' model as it is the simplest model that can capture bubble rebounds.

\subsection{Anisotropy parameter solutions}
Now that we have defined the anisotropy parameter and selected a collapse model, we will proceed to present three solution methods. We will start with analytic solutions, followed by semi-analytic and boundary element methods. All of these models rely on several key assumptions. The fluid is assumed to be incompressible, irrotational, and inviscid. Therefore, potential flow can be assumed. The bubble is considered to be spherically symmetrical with only the bubble radius varying with time.

\subsubsection{Analytic}
\label{sec:analytic_kelvin_impulse}
Much previous research has focused on calculating the Kelvin impulse analytically. \citet{Supponen2016} have presented anisotropy parameter solutions for several flow conditions. We include the solution for a flat plate here for completeness.

Treating the collapsing bubble as a sink in potential flow allows a solid boundary to be modelled using the method of images. This can then be solved for $\mathbf{F}'$ and the Kelvin impulse $\mathbf{I}$ \citep{Blake1988} yielding the equations

\begin{align}
\mathbf{F}' &= - \frac{\rho_L \mathbf{n}}{16 \pi Y^2} \\
\mathbf{I} &= -0.934 R_0^3 \sqrt{\Delta p \rho_L} \gamma^{-2} \mathbf{n} \label{eq:analytic_kelvin_impulse}
\end{align}
where $Y$ is the distance from the boundary; $\gamma = Y / R_0$ is the standoff distance; and $\Delta p$ is the initial difference between the pressure an infinite distance from the bubble and the internal pressure of the bubble, $\Delta p = p_\infty - p_V$.

Substituting Eq. (\ref{eq:analytic_kelvin_impulse}) into (\ref{eq:anisotropy_scaling}) yields
\begin{equation}
\boldsymbol{\zeta} = -0.195 \gamma^{-2} \mathbf{n}.
\label{eq:analytic_anisotropy}
\end{equation}

\subsubsection{Semi-analytic}
\label{sec:semi_analytic_kelvin_impulse}
The flow field of a set of geometries can also be solved with the method of images. In particular, these include bubbles in corner geometries \citep{PhysRevFluids.3.081601} and bubbles inside square and triangular prisms \citep{Molefe2019}. The triangular prisms are limited to the three tessellating cases: equilateral triangles, isosceles right triangles, and 30$^{\circ}$-60$^{\circ}$-90$^{\circ}$ triangles. Although an analytic description of the flow can be achieved, solving for the Kelvin impulse quickly becomes rather involved. Therefore, we numerically integrate the pressure over the boundary on a mesh in order to calculate the Kelvin impulse. This allows these analytic models to be extended numerically. The velocity at a point is

\begin{equation}
\nabla\phi|_j = \sum_{i=0}^M m_b \frac{(\mathbf{x}_j - \mathbf{x}_i)}{4 \pi |\mathbf{x}_j - \mathbf{x}_i|^3} = m_b \left[\sum_{i=0}^M \frac{(\mathbf{x}_j - \mathbf{x}_i)}{4 \pi |\mathbf{x}_j - \mathbf{x}_i|^3}\right] = m_b\nabla\phi_j'
\end{equation}
where $j$ is any point $j \neq i$ and $M$ is the total number of mirror sinks. $i = 0$ represents the bubble sink and $i > 0$ represent the mirror sinks. Although an infinite number of mirror sinks would be required to exactly match the boundary conditions, relatively few are required to adequately predict the anisotropy value.

Thus,
\begin{equation}
\nabla\phi_j' = \sum_{i=0}^M \frac{(\mathbf{x}_j - \mathbf{x}_i)}{4 \pi |\mathbf{x}_j - \mathbf{x}_i|^3}.
\label{eq:grad_phi_prime_sa}
\end{equation}

By assuming that the pressure is constant across an element in the boundary mesh, Eq. (\ref{eq:F_prime_definition}) can be discretized over the boundary elements to get
\begin{equation}
\mathbf{F}' = -\frac{1}{2}\rho \sum_{j=1}^N A_j |\nabla\phi_j'|^2 \mathbf{n}_j.
\label{eq:discretised_F_prime}
\end{equation}
where the points $j$ are the centroids of each boundary element, $A_j$ is the area of the boundary element at $j$, and $N$ is the total number of boundary elements.

Equation (\ref{eq:grad_phi_prime_sa}) is substituted into Eq. (\ref{eq:discretised_F_prime}) which is combined with Eq. (\ref{eq:anisotropy}) and solved numerically for the anisotropy parameter $\boldsymbol{\zeta}$.

\subsubsection{Boundary element method}
\label{sec:bem_kelvin_impulse}

We use the boundary element method to model geometries for which the flow field cannot be solved analytically. The boundary element method is a potential flow model that represents a boundary as a discretized distribution of potential flow elements. The full derivation and solution methodology used here is discussed in our previous work \cite{Andrews2020} and the main steps are summarized here. Each element $i$ of the boundary is represented as a single sink, positioned at the centroid of the element $\mathbf{x}_i$. The element is assumed to have a sink density $\sigma_i$ that is constant over its area $A_i$ such that the single sink has strength $\sigma_i A_i$. Thus, the velocity induced by an element is defined as
\begin{equation}
\nabla\phi|_j = \frac{\sigma_i A_i(\mathbf{x}_j - \mathbf{x}_i)}{4 \pi |\mathbf{x}_j - \mathbf{x}_i|^3},
\end{equation}
where $j$ is any point such that $\mathbf{x}_j \neq \mathbf{x}_i$.

Considering both the bubble and any boundaries, for any boundary element centroid point $j$, the velocity is given by the sum of the bubble sink and $N$ boundary element sinks
\begin{equation}
\nabla\phi|_j = \nabla\phi_b|_j + \nabla\phi_w|_j = \frac{m_b (\mathbf{x}_j - \mathbf{x}_b)}{4 \pi |\mathbf{x}_j - \mathbf{x}_b|^3} +  \sum_{i=1, i \neq j}^N m_b \frac{\sigma_i' A_i(\mathbf{x}_j - \mathbf{x}_i)}{4 \pi |\mathbf{x}_j - \mathbf{x}_i|^3} + \frac{m_b \sigma_j'}{2} \mathbf{n}_j.
\end{equation}
The summation of boundary element sinks is valid for any position $i \neq j$. When $i = j$ a singularity would be encountered. In the normal direction this is treated with the standard $\nabla\phi|_{j} \cdot \mathbf{n}_j = \sigma_j / 2$ \citep{Andrews2020}. An element is assumed to have no net effect on the tangential velocity at its centroid so no additional term is included. Note again that the boundary sink densities scale directly with the bubble sink strength such that $\sigma_j = m_b \sigma_j'$ where $\sigma_j'$ is the boundary sink strength densities computed for $m_b = 1$ m$^3$/s.

Thus, the velocity $\nabla\phi$ at any given bubble position, geometry, and time can be represented by a constant vector $\nabla\phi'$ multiplied by the bubble sink strength
\begin{equation}
\nabla\phi|_j = m_b \left[\frac{(\mathbf{x}_j - \mathbf{x}_b)}{4 \pi |\mathbf{x}_j - \mathbf{x}_b|^3} +  \sum_{i=1, i \neq j}^N \frac{\sigma_i' A_i(\mathbf{x}_j - \mathbf{x}_i)}{4 \pi |\mathbf{x}_j - \mathbf{x}_i|^3} + \frac{\sigma_j'}{2} \mathbf{n}\right] = m_b\nabla\phi_j',
\end{equation}
such that
\begin{equation}
\nabla\phi_j' = \frac{(\mathbf{x}_j - \mathbf{x}_b)}{4 \pi |\mathbf{x}_j - \mathbf{x}_b|^3} +  \sum_{i=1, i \neq j}^N \frac{\sigma_i' A_i(\mathbf{x}_j - \mathbf{x}_i)}{4 \pi |\mathbf{x}_j - \mathbf{x}_i|^3} + \frac{\sigma_j'}{2} \mathbf{n}.
\label{eq:grad_phi_prime_bem}
\end{equation}

As with the semi-analytic solution, Eq. (\ref{eq:F_prime_definition}) is discretized to produce Eq. (\ref{eq:discretised_F_prime}), which is combined with Eq. (\ref{eq:grad_phi_prime_bem}) to compute $\mathbf{F}'$. This $\mathbf{F}'$ is then substituted into Eq. (\ref{eq:anisotropy}) and solved numerically for the anisotropy parameter $\boldsymbol{\zeta}$.

\subsection{Experimental data and analysis}
\begin{figure}
\centerline{\includegraphics{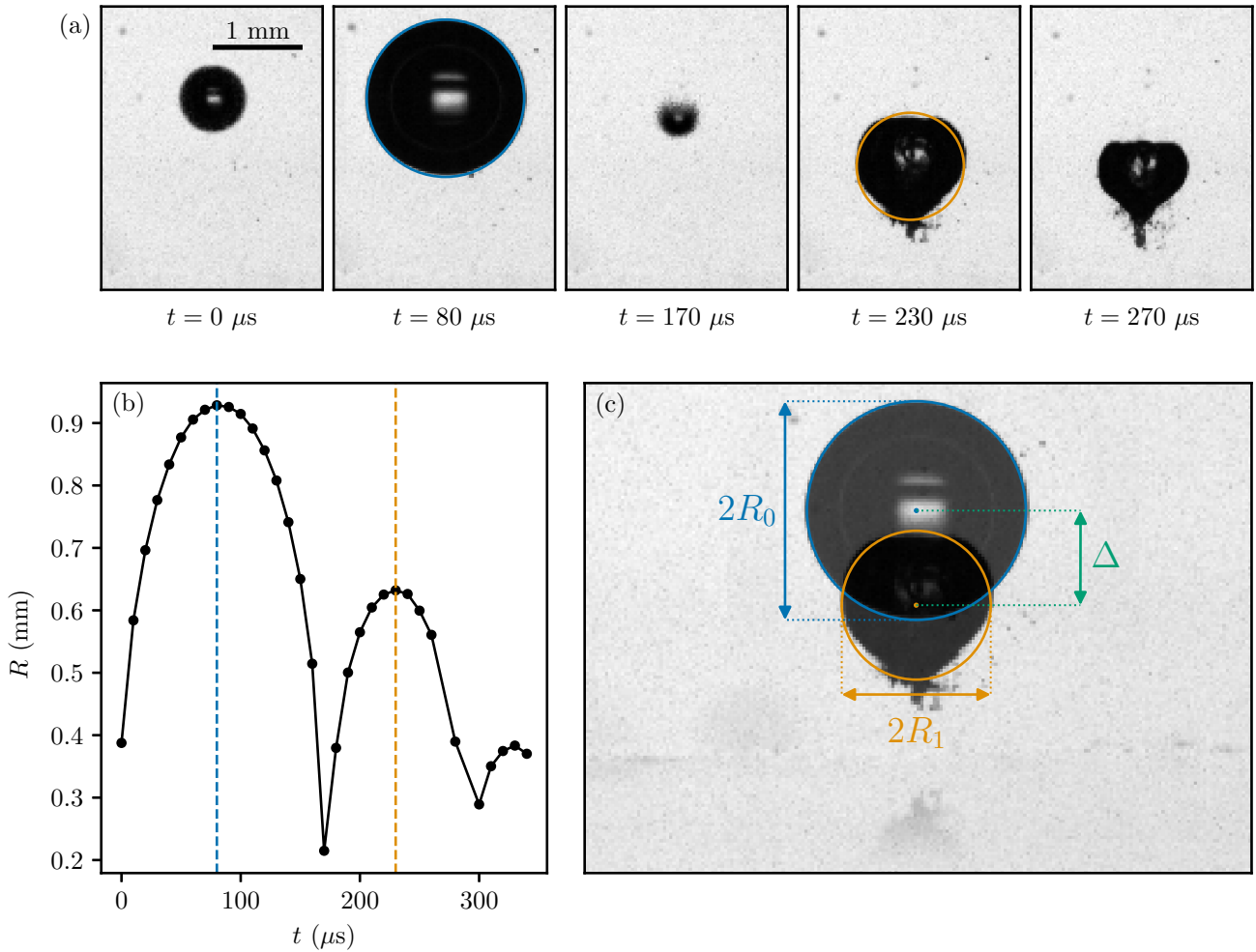}}
\caption{(a) Cropped frames of a bubble collapsing near a boundary. Circles with area equal to the bubble are shown for each of the size maxima ($t = 80 \mu$s and $t = 230 \mu$s). (b) Plot showing the measured bubble radius for the measured time span. (c) Composite of the two size maxima frames with equivalent circles and measured parameters annotated. $R_0$ is the initial maximum bubble radius. $R_1$ is the maximum bubble radius during the rebound. $\Delta$ is the displacement of the bubble centroid between the first and second bubble maxima.}
\label{fig:experiment_displacement}
\end{figure}

Bubble displacement is defined by its direction and distance. Previous work has characterized the bubble displacement direction in complex geometries \citep{PhysRevFluids.3.081601, Molefe2019, Andrews2020}. Here, we characterize the bubble displacement distance. The displacement distance $\Delta$ is measured from the position at the initial maximum size of the bubble and terminating at the position of the bubble at its maximum rebound size. These measurements are defined in Fig. \ref{fig:experiment_displacement}.

Some previous work has measured the bubble displacement from inception to end of the first collapse \citep{Supponen2016}. However, this measurement is difficult due to the very rapid growth, collapse, and movement of the bubble at these points in time. Conversely, at its maximum size, the bubble has a minimum rate of change of radius and minimum displacement velocity. Thus, measuring the bubble at the size maxima yields more robust measurements.

In addition to the bubble displacement, we measure the rebound size of the bubble $R_1$. In this work, the displacement and rebound size are both normalized by the maximum bubble radius $R_0$.

Experimental data used in this research are from prior investigations of jet direction for bubbles in corner geometries \citep{PhysRevFluids.3.081601, cornersdata}, inside triangular and square prisms \citep{Molefe2019, prismsdata}, and above slot geometries \citep{Andrews2020, slotsdata}. All of these experiments used laser-induced cavitation, with a microscope objective as the focusing optic. In addition, data has been gathered for a flat plate using the same experimental setup as \citet{Andrews2020}. Recordings were post-processed with Python, as in our previous work \citep{Andrews2020}. For each frame, the background was subtracted and a binary threshold filter was applied. The number of white pixels was taken as the bubble area $A$ and the centroid of these pixels as the bubble centroid. The radius was determined by assuming a spherical bubble such that $R = \sqrt{A / \pi}$. This assumption holds reasonably well for bubbles near their first and second size peaks as shown in Fig. \ref{fig:experiment_displacement}. When the bubble is not spherical, this can simply be taken as an equivalent bubble radius.

\section{Results and discussion}
In this section we apply our numerical framework to a series of geometries with varying complexity. We begin by comparing models, we then generate anisotropy magnitude maps for four complex geometries, and finally show that experimental data collapses when combined with anisotropy parameter predictions.
\subsection{Numerical anisotropy parameter calculations}
\label{sec:numerical_results}
\subsubsection{Flat plate model comparison}
We start with a flat plate geometry, which can be treated with all three models: analytic, semi-analytic, and boundary element method. The analytic solution is $\zeta = 0.195 \gamma^{-2}$ which is simply the magnitude of Eq. (\ref{eq:analytic_anisotropy}). The semi-analytic solution uses a mirror sink and integrates the pressure over the boundary on the mesh of a $1$ m$^2$ plate. The boundary element method solution uses the same mesh as the semi-analytic solution to compute boundary sink strengths and the pressure integration. In this comparison the mesh consisted of 18~020 elements with most elements concentrated in the central region. Element lengths ranged between 0.4 mm and 13.9 mm. The bubbles were positioned at $Y = 5$ mm from the boundary and the bubble radius $R_0$ was varied between $0.5$ mm and $5$ mm to produce a range of standoff distances between $1$ and $10$.

\begin{figure}
\centerline{\includegraphics{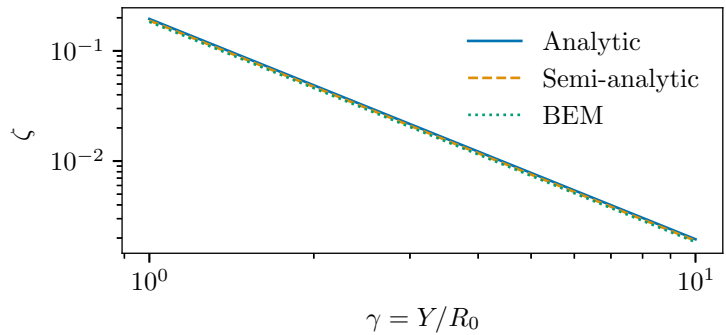}}
\caption{The anisotropy parameter $\zeta$ against standoff $\gamma$ for a simple flat plate, compared between the analytic, semi-analytic, and boundary element method models, showing a near-perfect match.}
\label{fig:flat_plate_model_comparison}
\end{figure}

As $\mathbf{F}'$ depends only on bubble position, the series of bubbles can be characterized by a single $\mathbf{F}'$ value. The semi-analytic model produced an $\mathbf{F}'$ value $0.5 \%$ lower than the analytic model and the boundary element method model produced an $\mathbf{F}'$ value $3.2 \%$ lower than the analytic model. As is evident from Eq. (\ref{eq:anisotropy}), these cause a proportional decrease in the measured anisotropy magnitude. Figure \ref{fig:flat_plate_model_comparison} shows the comparison between the anisotropy magnitude values calculated by these models. Both the semi-analytic and boundary element method solutions predict lower anisotropy magnitude values than the analytic solution, this is in part due to integrating the pressure over a finite plate rather than the infinite plate assumed by the analytic solution. There is also some difference due to the discretization of the boundary. For the semi-analytic solution the discretization is only for the integration of pressure over the boundary, while for the boundary element method solution the discretization is in boundary conditions, flow solution, and also the integration of pressure. Finally, there is some difference due to the Rayleigh-Plesset model used. The analytic solution uses a `\ref{eq:rp-no-internal-gas}' type model whereas the other two solutions use the `\ref{eq:rp-inertial}' model. Despite these differences, the three models produce very consistent anisotropy parameter values, and we conclude that the model is insensitive to these details.

\subsubsection{Anisotropy maps for complex geometries}
Unlike the analytic and semi-analytic models, the boundary element method can be applied to any geometry. In addition, the most expensive part of the boundary element method only needs to be calculated once for any geometry and the result can be used for all bubble positions. However, it is vulnerable to ill-conditioned geometries. Although this is generally unlikely to occur, it is a significant problem for very regular, highly enclosed geometries such as the prisms presented by \citet{Molefe2019}. In contrast, the semi-analytic method can operate with far fewer sinks and is stable for all valid geometries, but the full calculation must be performed for every bubble position.

The primary computational limitation of the boundary element method is the memory usage. Memory usage scales with $N^2$ where $N$ is the number of boundary elements. The semi-analytic method separates this scaling into two parts: $N$ points on the pressure mesh and $M$ mirror sinks. The memory thus scales with $N M$. Where the semi-analytic model is possible, far fewer sinks $M$ are required to model the boundary compared to the boundary element method, so far more sinks $N$ can be used to model the pressure mesh within the same memory constraint. This allows for the pressure to be more precisely resolved.

\begin{figure}
\centerline{\includegraphics{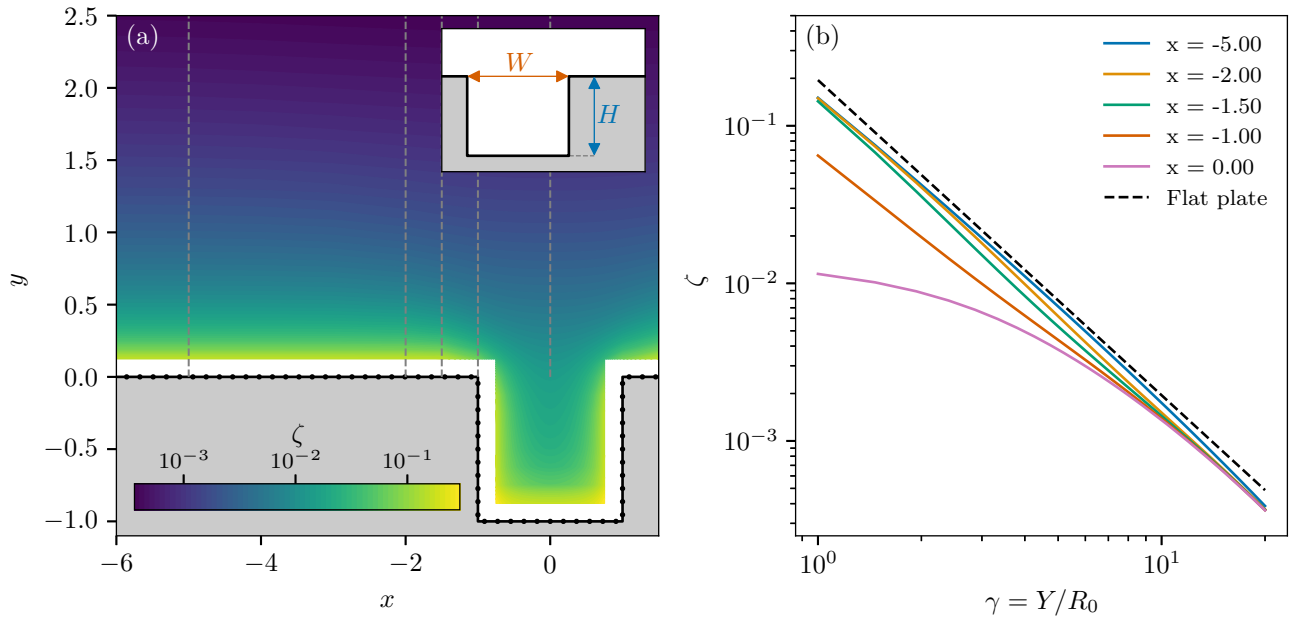}}
\caption{(a) A contour plot of anisotropy magnitude for bubbles positioned around a slot with $H / W = 1$, $W / R_0 =  8$ and $R_0=0.5~\rm{mm}$. $x = 2X / W$ and $y = Y / W$ where $X$ is the horizontal position of the bubble from the slot center, $Y$ is the vertical position of the bubble from the upper surface, and $W$ is the slot width \citep{Andrews2020}. The slot boundary is indicated in solid black with boundary sink positions displayed as black dots. Gray dashed lines correspond to the horizontal positions in (b). (b) Anisotropy magnitude against standoff distance from the upper surface of the slot for five horizontal positions. The dashed line is the analytic solution for a flat plate.}
\label{fig:slot_anisotropy_contour}
\end{figure}

Here we present anisotropy magnitude maps for a selection of complex geometries. Figure \ref{fig:slot_anisotropy_contour}(a) shows an anisotropy magnitude map for a slot geometry of the type treated by \citet{Andrews2020}. The slot has a width and height $W = H = 8 R_0$ for $R_0 = 0.5$ mm. $N = 22~939$ elements were used to resolve the boundary. In contrast to the previous work, the bubble size here is important as it is required for computing the anisotropy magnitude. Due to the finite size of the bubble, anisotropy values extremely close to the boundary do not have a clear physical meaning despite being numerically feasible. Thus, the white area near the boundary in Fig. \ref{fig:slot_anisotropy_contour}(a) is an area of width $R_0$ for which the anisotropy magnitude was not computed.

At a large horizontal distance from the center of the slot, corresponding to the most negative $x$ values in Fig. \ref{fig:slot_anisotropy_contour}(b), the boundary becomes similar to a flat plate, with the anisotropy magnitude depending primarily on the vertical standoff distance $\gamma = Y / R_0$. As the bubble horizontally approaches the slot, the slot causes a general reduction in anisotropy due to the increase of fluid volume below the bubble. At the center of the slot, the anisotropy contributions from the slot sides cancel out horizontally due to symmetry. The anisotropy is found to be strongest in the bottom corners of the slot where the bubble is most confined on one side but not the other.

Figure \ref{fig:slot_anisotropy_contour}(b) shows anisotropy magnitude curves for five different horizontal positions with a range of standoff values where the standoff $\gamma = Y / R_0$. These are compared to the anisotropy magnitude for a flat plate. Far from the slot horizontally, the $x = -5$ curve is very close to the flat plate curve as the effect of the slot is minimal. For horizontal positions closer to the slot, these curves deviate significantly from the flat plate curve. 

\begin{figure}
\centerline{\includegraphics{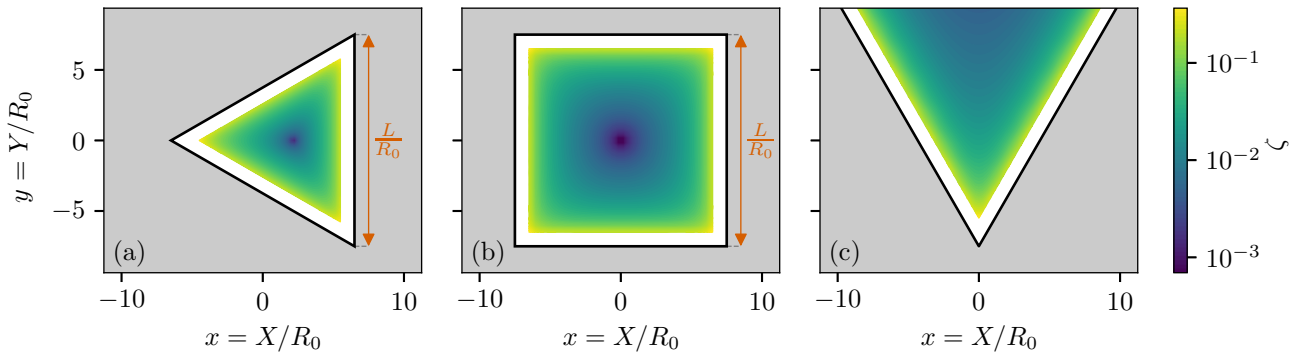}}
\caption{(a) A contour plot of anisotropy magnitude for bubbles positioned in an equilateral triangular prism. $L / R_0 = 15$ where $L$ is the side length of the triangle. (b) A contour plot of anisotropy magnitude for bubbles positioned in a square prism. $L / R_0 = 15$ where $L$ is the side length of the square. (c) A contour plot of anisotropy magnitude for bubbles positioned in a corner with angle $\theta_c = \pi / 3$. These geometries are of the type investigated by \citet{PhysRevFluids.3.081601} and \citet{Molefe2019}. The boundaries are indicated by solid black lines. The area outside of the fluid domain is shaded gray.}
\label{fig:combined_anisotropy_contours}
\end{figure}

The anisotropy magnitude for triangular prisms, square prisms, and corner geometries were calculated using the semi-analytic method. Figure \ref{fig:combined_anisotropy_contours} shows contour plots for anisotropy magnitude in triangular (Fig. \ref{fig:combined_anisotropy_contours}(a)) and square (Fig. \ref{fig:combined_anisotropy_contours}(b)) prisms of the type treated by \citet{Molefe2019} and in a corner of the type treated by \citet{PhysRevFluids.3.081601} (Fig. \ref{fig:combined_anisotropy_contours}(c)). For the same reason as given for the slot geometries, anisotropy was not calculated in the white areas near the boundaries. The two prisms have side length $L = 15 R_0$. The triangular prism used $M = 12~675$ image sinks and the square prism used $M = 4225$ image sinks, both with $N \approx 20~000$ elements in the pressure mesh. The corner has an internal angle $\theta_c = \pi / 3$ and thus used $M = 5$ mirror sinks. Here the anisotropy is at a maximum near the corners of each shape showing similar tendencies as the slot geometry. Towards the center of the shape, bubbles experience decreasing anisotropy as the bubble collapse is less impeded by the boundaries. At the exact center of the prisms, the anisotropy is expected to be zero due to symmetry.

\subsection{Experimental results}
It has previously been shown that bubble displacement and rebound size depend on the anisotropy parameter \citep{Supponen2016, Supponen2018}. In this section we present experimental measurements of displacement and rebound size and compare them to anisotropy parameter predictions for each collapse event.
\label{sec:experimental_results}
\subsubsection{Data post-processing}
\begin{figure}[htbp]
    \centering
    \includegraphics{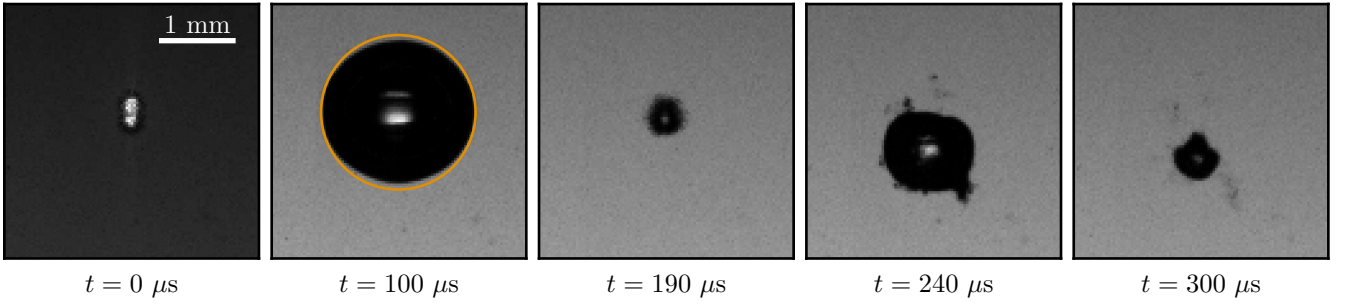}
    \caption{A series of frames showing a bubble collapsing near a slot. The laser enters from the top of frame. The frame at $t = 0~\mu$s shows the initial plasma generated by the laser. The orange line on the frame at $t = 100~\mu$s shows a perfect circle encompassing the bubble for comparison. The eccentricity of the bubble was 0.36 at $t = 100~\mu$s.}
    \label{fig:elongated_plasma}
\end{figure}

The models presented in this work assume a spherical bubble. However, bubbles in experiments deviate from being perfectly spherical to varying degrees. When a bubble is initially nucleated a plasma is formed around the point at which the laser is focused. If the angle of convergence of the beam is too small (having a low equivalent numerical aperture), the energy density of the beam can remain high away from the laser focus. This leads to the plasma forming in an elongated shape, and sometimes even forming multiple separate spots of plasma \citep{Tagawa2016}. This elongation leads to an oblate bubble at the maximum bubble size and increases the spread of data for measured displacement. Figure \ref{fig:elongated_plasma} demonstrates this occurrence. In the first frame, at $t = 0~\mu$s, the plasma is visible and is elongated to the point of having two almost-separated sections. In the second frame, at $t = 100~\mu$s where the bubble is at its maximum size, the bubble is slightly oblate. We quantify the deviation from a perfect sphere using the eccentricity of the bubble image, where eccentricity is defined as the eccentricity of an ellipse with second-moments equal to the bubble image region. At its maximum size, the bubble in Fig. \ref{fig:elongated_plasma} had a measured eccentricity of 0.36. By contrast, the bubble shown in Fig. \ref{fig:experiment_displacement} had an eccentricity of 0.25 at its maximum size. The third frame, at $t = 190~\mu$s, shows that the bubble collapses as expected. However, when it rebounds in the fourth frame at $t = 240~\mu$s, it retains the oblate deformation.

\begin{figure}[htbp]
    \centering
    \includegraphics{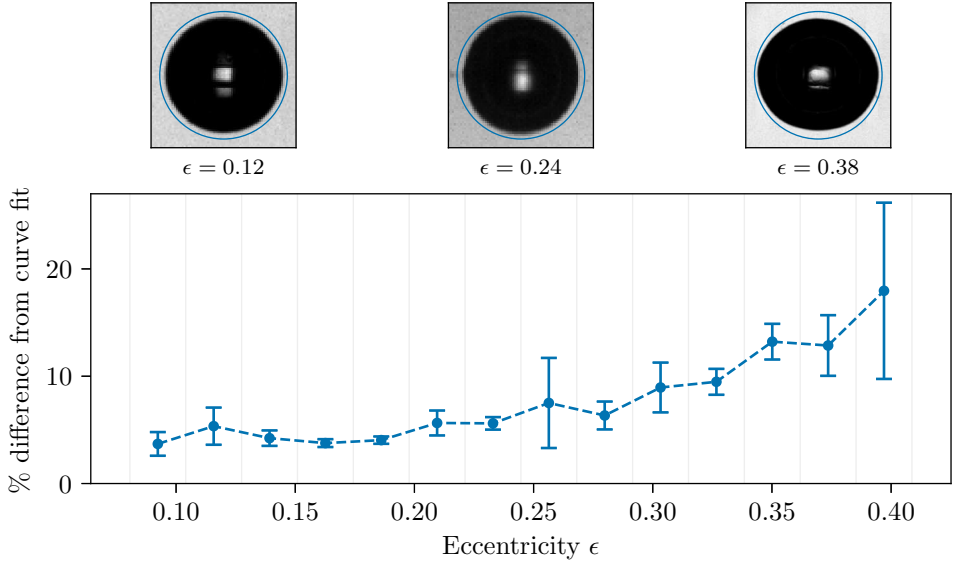}
    \caption{A plot showing how the spread of data varies with eccentricity. The spread is defined as the percentage difference between a data point and a curve fitted across all data. Data are placed in discrete bins and the mean of each bin is shown at the bin center with the 95\% confidence interval shown by the error bars. Only bins with five or more data points are included. Frames showing examples of select eccentricities are displayed above the plot with perfect circles surrounding the bubbles for comparison.}
    \label{fig:spread_vs_eccentricity}
\end{figure}

Bubbles that are deformed upon formation are expected to result in spread in experimental measurements. Figure \ref{fig:spread_vs_eccentricity} shows the spread of data as a function of bubble eccentricity at their maximum size. Here the spread is defined as the percentage difference between a data point and a power law curve that has been fitted across all data. The curve fit used for this determination is shown in Fig. \ref{fig:all_data_vs_anisotropy}(a). Although the error bars are quite large in some areas of Fig. \ref{fig:spread_vs_eccentricity}, it is clear that greater eccentricity leads to a greater spread in measured displacement. In order to compare consistent bubbles, and because we expect non-spherical bubbles to deviate most strongly from our model, we filter our experimental data by eccentricity. Thus, by filtering out highly eccentric bubbles, the spread of data can be reduced to more closely align with the idealized scenario of a spherical bubble. Filtered data are shown in Fig. \ref{fig:all_data_vs_anisotropy} with unfiltered data shown in the inset plots. As expected, when the data are filtered, the spread of data reduces and many of the most significant outliers are removed. The eccentricity values used to filter these data are set separately for each geometry. Some geometries have many data points (the slots data set has 5094 points) whereas others have far fewer (the flat plate data set has only 82). Thus, much more stringent eccentricity filters are applied to larger data sets. Overall, with eccentricity limits varying between 0.2 (for slots) and 0.28 (for the flat plate), 2124 data points were retained after filtering across all data, comprising 31\% of the available data.

\subsubsection{Experimental data for all geometries}
\begin{figure}
\centerline{\includegraphics{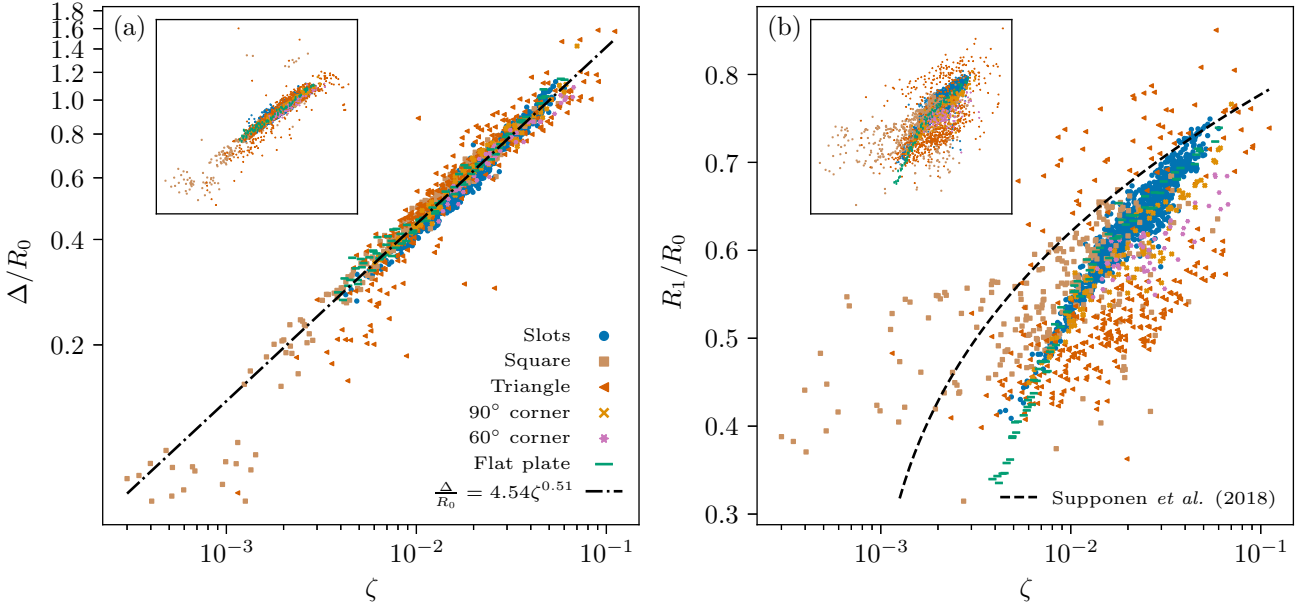}}
\caption{Experimental data for a flat plate, a series of slots, a triangular prism, a square prism, and two corners. Experimental data is plotted against anisotropy magnitude predictions using the numerical models in this work. (a) Normalized bubble displacement $\Delta / R_0$ against anisotropy magnitude $\zeta$. A power law curve fit for all the data combined is shown by the black dash-dotted line. (b) Ratio of second bubble size maximum to first bubble size maximum $R_1 / R_0$ against anisotropy magnitude $\zeta$. The black dashed line is derived from \citet{Supponen2018}. Inset axes of both plots include all data unfiltered.}
\label{fig:all_data_vs_anisotropy}
\end{figure}

For any geometry, the displacement and radius ratio measurements can be plotted against predicted anisotropy magnitudes. All considered geometries are shown in Fig. \ref{fig:all_data_vs_anisotropy}. The corner, square prism and triangular prism data are plotted with semi-analytic anisotropy model predictions. The corner used $M = 3$ or $M = 5$, depending on corner angle, and $N \approx 50~000$ elements in the pressure mesh. The triangular prism model used $M = 11~163$ image sinks with $N \approx 30~000$ elements in the pressure mesh. The square prism model used $M = 4225$ image sinks with $N \approx 50~000$ elements in the pressure mesh. The flat plate and slot data are plotted using boundary element method anisotropy model predictions with $N = 19~896$ and $19~662 \leq N \leq 20~142$, respectively. There is some variation in the number of elements used for the slot geometries due to varying slot sizes.

For the bubble displacement, shown in Fig. \ref{fig:all_data_vs_anisotropy}(a), all data collapses well onto a single line within the experimental variance of the data. The most significant deviation from this curve is at very low anisotropy magnitude values, where the bubbles move very little. Because of the small displacement, random noise is expected to contribute more to the deviation, although the data appears biased towards smaller displacement values. However, data in this region is sparse, so care should be taken to draw any conclusions from it.
We further observe deviations at very high anisotropy magnitude values, where bubbles are very close to the walls and thus can become constrained. This constraint limits any meaningful measurement of the displacement and therefore also precludes any validation of the numerical model in this regime.
The collapsed curve conforms to the power law $\Delta / R_0 = 4.54 \zeta^{0.51}$.

Figure \ref{fig:all_data_vs_anisotropy}(b) shows the radius ratio $R_1/R_0$ as a function of the predicted anisotropy magnitude $\zeta$. It is clear that there is much more spread in the data compared to the displacement data of Fig. \ref{fig:all_data_vs_anisotropy}(a). However, looking first at the flat plate and slot data, both data sets approximately collapse onto a single curve.

The data for the two corners has a slightly higher spread than the slots data, and tends to have a lower radius ratio at high anisotropies. Finally, the triangle and square data have the highest spread. The biggest grouping of square data falls close to the collapsed slot and flat plate data, however the triangle data is most significantly grouped below the other data sets. Due to experimental limitations, the triangle and square data is comprised of relatively small bubbles at a greater distance from the camera, producing small images in the frame with short collapse times. Smaller images cannot be measured as accurately as larger images which leads to greater measurement error in determining the radius ratio. In addition, shorter collapse times lead to fewer frames of data which can cause the radius to be measured at the wrong time. However, neither of these factors can fully account for such spread of data. It is noted that more confined geometries tend to have higher spreads of radius ratio which suggests that shockwave reflections might play a more important role in such geometries. Despite these variations, much of the data collapses reasonably well (in particular the slots and flat plate data), suggesting that it does strongly depend on the anisotropy parameter as has been previously reported \citep{Supponen2018}.

\citet{Supponen2018} presented the ratio of energy between first and second size maxima as a function of the anisotropy parameter. The energy was calculated from the radius and thus the radius ratio function can be determined from the given rebound energy ratio function. This function is plotted alongside the data in Fig. \ref{fig:all_data_vs_anisotropy}(b) and significantly deviates from the experimental data. Although the cause of this deviation is not known, we speculate that minor variations in the experimental methodology could lead to different amounts of energy dissipation during the collapse, thus producing different radius ratio curves. The sensitivity to experimental conditions could also explain the large spread in the data in Fig. \ref{fig:all_data_vs_anisotropy}(b). One such parameter is the amount of non-condensable gas present in the bubble which is difficult to measure experimentally and has previously been shown to strongly affect the size of bubble rebounds \citep{Tinguely2012, Trummler2021}. Variation in the amount of non-condensable gas is therefore a good candidate as the cause of the variation between previous work and the data presented here. This variation of rebound radius presents a challenge in the use of the collapsed curves presented in the current work. Displacement measured from bubble inception to the end of the first collapse \citep{Supponen2016} is unaffected by rebound size and so the relation between displacement and the anisotropy parameter would be expected to remain constant across all experiments. However, displacement measured from the initial bubble size peak to the rebound size peak would vary as the rebound size varies. Thus, it is expected that the collapsed curves presented in Fig. \ref{fig:all_data_vs_anisotropy} are not universal.

Despite the lack of universality, this data corroborates the assertion of \citet{Supponen2016} that many bubble collapse properties depend primarily on the anisotropy parameter. The collapse of the data also serves to validate the numerical models presented in this work. These models, in combination with established scaling functions \citep{Supponen2016}, can therefore be used to predict bubble collapse properties, such as jet velocity and jet volume, for any rigid geometry.

\section{Conclusion and outlook}
In this work we have presented a numerical model for computing the anisotropy parameter defined by \citet{Supponen2016}. We have applied this model to a series of complex geometries and demonstrated how the anisotropy parameter varies with bubble position. The anisotropy magnitude is found to be highest when the bubble is highly confined in one or more directions but with open fluid in another direction. This configuration is mostly found near concave corners of geometries. The anisotropy magnitude is found to be at a minimum far from geometries and in highly symmetric areas.

We have experimentally measured two parameters of bubble collapse that can both be measured robustly: the bubble displacement and ratio of bubble radius between the first two size maxima. Using the anisotropy magnitude predicted by the numerical model, we have shown that the bubble displacement collapses onto a single curve which conforms to the power law $\Delta / R_0 = 4.54 \zeta^{0.51}$ where $\Delta / R_0$ is the dimensionless displacement and $\zeta$ is the anisotropy parameter magnitude. A large portion of the radius ratio data collapses approximately onto a single curve, however there is very high spread in some data sets and some significant overall deviation from the collapsed curve. In addition, it is clear that this data does not collapse onto the same curve as was presented by \citet{Supponen2018}. We suggest that this variation could be significantly influenced by experimental differences such as the non-condensable gas content of the bubble. Such a variation should also lead to a variation in shockwave energy so the energy partition could be used to test this theory \citep{Tinguely2012}. We recommend further investigation of this discrepancy, aiming to identify the key parameters that lead to such variation between experimental setups and how those parameters can be controlled.

Previous research has shown that the anisotropy parameter is a good predictor of various collapse properties in a range of simple geometries \citep{Supponen2016}. The anisotropy parameter model presented here, in combination with experimental data, has shown that the anisotropy parameter remains a good predictor of collapse properties, even in complex geometries. Thus, this model can be combined with scaling laws, such as those presented by \citet{Supponen2016}, to predict many bubble collapse properties, such as jet velocity, jet volume, and jet impact time. The anisotropy vector direction produced by this model is equivalent to the directions produced by models that have already been investigated in connection with the bubble displacement direction \citep{PhysRevFluids.3.081601, Molefe2019, Andrews2020}. This model can therefore provide a complete prediction of the bubble displacement for any rigid geometry.

\begin{acknowledgements}
We thank Lebo Molefe for insightful discussion and contribution of data used in this research. We acknowledge financial support from the EPSRC under Grant No. EP/P012981/1. 
\end{acknowledgements}


\begin{thebibliography}{53}%
\makeatletter
\providecommand \@ifxundefined [1]{%
 \@ifx{#1\undefined}
}%
\providecommand \@ifnum [1]{%
 \ifnum #1\expandafter \@firstoftwo
 \else \expandafter \@secondoftwo
 \fi
}%
\providecommand \@ifx [1]{%
 \ifx #1\expandafter \@firstoftwo
 \else \expandafter \@secondoftwo
 \fi
}%
\providecommand \natexlab [1]{#1}%
\providecommand \enquote  [1]{``#1''}%
\providecommand \bibnamefont  [1]{#1}%
\providecommand \bibfnamefont [1]{#1}%
\providecommand \citenamefont [1]{#1}%
\providecommand \href@noop [0]{\@secondoftwo}%
\providecommand \href [0]{\begingroup \@sanitize@url \@href}%
\providecommand \@href[1]{\@@startlink{#1}\@@href}%
\providecommand \@@href[1]{\endgroup#1\@@endlink}%
\providecommand \@sanitize@url [0]{\catcode `\\12\catcode `\$12\catcode
  `\&12\catcode `\#12\catcode `\^12\catcode `\_12\catcode `\%12\relax}%
\providecommand \@@startlink[1]{}%
\providecommand \@@endlink[0]{}%
\providecommand \url  [0]{\begingroup\@sanitize@url \@url }%
\providecommand \@url [1]{\endgroup\@href {#1}{\urlprefix }}%
\providecommand \urlprefix  [0]{URL }%
\providecommand \Eprint [0]{\href }%
\providecommand \doibase [0]{https://doi.org/}%
\providecommand \selectlanguage [0]{\@gobble}%
\providecommand \bibinfo  [0]{\@secondoftwo}%
\providecommand \bibfield  [0]{\@secondoftwo}%
\providecommand \translation [1]{[#1]}%
\providecommand \BibitemOpen [0]{}%
\providecommand \bibitemStop [0]{}%
\providecommand \bibitemNoStop [0]{.\EOS\space}%
\providecommand \EOS [0]{\spacefactor3000\relax}%
\providecommand \BibitemShut  [1]{\csname bibitem#1\endcsname}%
\let\auto@bib@innerbib\@empty
\bibitem [{\citenamefont {Rayleigh}(1917)}]{Rayleigh1917}%
  \BibitemOpen
  \bibfield  {author} {\bibinfo {author} {\bibfnamefont {L.}~\bibnamefont
  {Rayleigh}},\ }\bibfield  {title} {\bibinfo {title} {{VIII. On the pressure
  developed in a liquid during the collapse of a spherical cavity}},\ }\href
  {https://doi.org/10.1080/14786440808635681} {\bibfield  {journal} {\bibinfo
  {journal} {The London, Edinburgh, and Dublin Philosophical Magazine and
  Journal of Science}\ }\textbf {\bibinfo {volume} {34}},\ \bibinfo {pages}
  {94} (\bibinfo {year} {1917})}\BibitemShut {NoStop}%
\bibitem [{\citenamefont {Luo}\ \emph {et~al.}(2016)\citenamefont {Luo},
  \citenamefont {Ji},\ and\ \citenamefont {Tsujimoto}}]{Luo2016}%
  \BibitemOpen
  \bibfield  {author} {\bibinfo {author} {\bibfnamefont {X.~W.}\ \bibnamefont
  {Luo}}, \bibinfo {author} {\bibfnamefont {B.}~\bibnamefont {Ji}},\ and\
  \bibinfo {author} {\bibfnamefont {Y.}~\bibnamefont {Tsujimoto}},\ }\bibfield
  {title} {\bibinfo {title} {{A review of cavitation in hydraulic machinery}},\
  }\href {https://doi.org/10.1016/S1001-6058(16)60638-8} {\bibfield  {journal}
  {\bibinfo  {journal} {Journal of Hydrodynamics}\ }\textbf {\bibinfo {volume}
  {28}},\ \bibinfo {pages} {335} (\bibinfo {year} {2016})}\BibitemShut
  {NoStop}%
\bibitem [{\citenamefont {Lu}\ \emph {et~al.}(2016)\citenamefont {Lu},
  \citenamefont {Yuan}, \citenamefont {Luo}, \citenamefont {Yuan},
  \citenamefont {Zhou},\ and\ \citenamefont {Sun}}]{Lu2016}%
  \BibitemOpen
  \bibfield  {author} {\bibinfo {author} {\bibfnamefont {J.}~\bibnamefont
  {Lu}}, \bibinfo {author} {\bibfnamefont {S.}~\bibnamefont {Yuan}}, \bibinfo
  {author} {\bibfnamefont {Y.}~\bibnamefont {Luo}}, \bibinfo {author}
  {\bibfnamefont {J.}~\bibnamefont {Yuan}}, \bibinfo {author} {\bibfnamefont
  {B.}~\bibnamefont {Zhou}},\ and\ \bibinfo {author} {\bibfnamefont
  {H.}~\bibnamefont {Sun}},\ }\bibfield  {title} {\bibinfo {title} {{Numerical
  and experimental investigation on the development of cavitation in a
  centrifugal pump}},\ }\href {https://doi.org/10.1177/0954408914557877}
  {\bibfield  {journal} {\bibinfo  {journal} {Proceedings of the Institution of
  Mechanical Engineers, Part E: Journal of Process Mechanical Engineering}\
  }\textbf {\bibinfo {volume} {230}},\ \bibinfo {pages} {171} (\bibinfo {year}
  {2016})}\BibitemShut {NoStop}%
\bibitem [{\citenamefont {Sagar}\ and\ \citenamefont
  {el~Moctar}(2020)}]{Sagar2020}%
  \BibitemOpen
  \bibfield  {author} {\bibinfo {author} {\bibfnamefont {H.~J.}\ \bibnamefont
  {Sagar}}\ and\ \bibinfo {author} {\bibfnamefont {O.}~\bibnamefont
  {el~Moctar}},\ }\bibfield  {title} {\bibinfo {title} {{Dynamics of a
  cavitation bubble near a solid surface and the induced damage}},\ }\bibfield
  {journal} {\bibinfo  {journal} {Journal of Fluids and Structures}\ }\href
  {https://doi.org/10.1016/j.jfluidstructs.2019.102799}
  {10.1016/j.jfluidstructs.2019.102799} (\bibinfo {year} {2020})\BibitemShut
  {NoStop}%
\bibitem [{\citenamefont {Ohl}\ \emph {et~al.}(2006)\citenamefont {Ohl},
  \citenamefont {Arora}, \citenamefont {Dijkink}, \citenamefont {Janve},\ and\
  \citenamefont {Lohse}}]{Ohl2006}%
  \BibitemOpen
  \bibfield  {author} {\bibinfo {author} {\bibfnamefont {C.~D.}\ \bibnamefont
  {Ohl}}, \bibinfo {author} {\bibfnamefont {M.}~\bibnamefont {Arora}}, \bibinfo
  {author} {\bibfnamefont {R.}~\bibnamefont {Dijkink}}, \bibinfo {author}
  {\bibfnamefont {V.}~\bibnamefont {Janve}},\ and\ \bibinfo {author}
  {\bibfnamefont {D.}~\bibnamefont {Lohse}},\ }\bibfield  {title} {\bibinfo
  {title} {{Surface cleaning from laser-induced cavitation bubbles}},\
  }\bibfield  {journal} {\bibinfo  {journal} {Applied Physics Letters}\
  }\textbf {\bibinfo {volume} {89}},\ \href {https://doi.org/10.1063/1.2337506}
  {10.1063/1.2337506} (\bibinfo {year} {2006})\BibitemShut {NoStop}%
\bibitem [{\citenamefont {Verhaagen}\ and\ \citenamefont {{Fern{\'{a}}ndez
  Rivas}}(2016)}]{Verhaagen2016}%
  \BibitemOpen
  \bibfield  {author} {\bibinfo {author} {\bibfnamefont {B.}~\bibnamefont
  {Verhaagen}}\ and\ \bibinfo {author} {\bibfnamefont {D.}~\bibnamefont
  {{Fern{\'{a}}ndez Rivas}}},\ }\bibfield  {title} {\bibinfo {title}
  {{Measuring cavitation and its cleaning effect}},\ }\href
  {https://doi.org/10.1016/j.ultsonch.2015.03.009} {\bibfield  {journal}
  {\bibinfo  {journal} {Ultrasonics Sonochemistry}\ }\textbf {\bibinfo {volume}
  {29}},\ \bibinfo {pages} {619} (\bibinfo {year} {2016})}\BibitemShut
  {NoStop}%
\bibitem [{\citenamefont {Birkin}\ \emph {et~al.}(2016)\citenamefont {Birkin},
  \citenamefont {Offin},\ and\ \citenamefont {Leighton}}]{Birkin2016}%
  \BibitemOpen
  \bibfield  {author} {\bibinfo {author} {\bibfnamefont {P.~R.}\ \bibnamefont
  {Birkin}}, \bibinfo {author} {\bibfnamefont {D.~G.}\ \bibnamefont {Offin}},\
  and\ \bibinfo {author} {\bibfnamefont {T.~G.}\ \bibnamefont {Leighton}},\
  }\bibfield  {title} {\bibinfo {title} {{An activated fluid stream - New
  techniques for cold water cleaning}},\ }\href
  {https://doi.org/10.1016/j.ultsonch.2015.10.001} {\bibfield  {journal}
  {\bibinfo  {journal} {Ultrasonics Sonochemistry}\ }\textbf {\bibinfo {volume}
  {29}},\ \bibinfo {pages} {612} (\bibinfo {year} {2016})}\BibitemShut
  {NoStop}%
\bibitem [{\citenamefont {Chahine}\ \emph {et~al.}(2016)\citenamefont
  {Chahine}, \citenamefont {Kapahi}, \citenamefont {Choi},\ and\ \citenamefont
  {Hsiao}}]{Chahine2016}%
  \BibitemOpen
  \bibfield  {author} {\bibinfo {author} {\bibfnamefont {G.~L.}\ \bibnamefont
  {Chahine}}, \bibinfo {author} {\bibfnamefont {A.}~\bibnamefont {Kapahi}},
  \bibinfo {author} {\bibfnamefont {J.~K.}\ \bibnamefont {Choi}},\ and\
  \bibinfo {author} {\bibfnamefont {C.~T.}\ \bibnamefont {Hsiao}},\ }\bibfield
  {title} {\bibinfo {title} {{Modeling of surface cleaning by cavitation bubble
  dynamics and collapse}},\ }\bibfield  {journal} {\bibinfo  {journal}
  {Ultrasonics Sonochemistry}\ }\href
  {https://doi.org/10.1016/j.ultsonch.2015.04.026}
  {10.1016/j.ultsonch.2015.04.026} (\bibinfo {year} {2016})\BibitemShut
  {NoStop}%
\bibitem [{\citenamefont {Verhaagen}\ \emph {et~al.}(2016)\citenamefont
  {Verhaagen}, \citenamefont {Zanderink},\ and\ \citenamefont {{Fernandez
  Rivas}}}]{Verhaagen2016a}%
  \BibitemOpen
  \bibfield  {author} {\bibinfo {author} {\bibfnamefont {B.}~\bibnamefont
  {Verhaagen}}, \bibinfo {author} {\bibfnamefont {T.}~\bibnamefont
  {Zanderink}},\ and\ \bibinfo {author} {\bibfnamefont {D.}~\bibnamefont
  {{Fernandez Rivas}}},\ }\bibfield  {title} {\bibinfo {title} {{Ultrasonic
  cleaning of 3D printed objects and Cleaning Challenge Devices}},\ }\href
  {https://doi.org/10.1016/j.apacoust.2015.06.010} {\bibfield  {journal}
  {\bibinfo  {journal} {Applied Acoustics}\ }\textbf {\bibinfo {volume}
  {103}},\ \bibinfo {pages} {172} (\bibinfo {year} {2016})}\BibitemShut
  {NoStop}%
\bibitem [{\citenamefont {Reuter}\ \emph {et~al.}(2017)\citenamefont {Reuter},
  \citenamefont {Lauterborn}, \citenamefont {Mettin},\ and\ \citenamefont
  {Lauterborn}}]{Reuter2017}%
  \BibitemOpen
  \bibfield  {author} {\bibinfo {author} {\bibfnamefont {F.}~\bibnamefont
  {Reuter}}, \bibinfo {author} {\bibfnamefont {S.}~\bibnamefont {Lauterborn}},
  \bibinfo {author} {\bibfnamefont {R.}~\bibnamefont {Mettin}},\ and\ \bibinfo
  {author} {\bibfnamefont {W.}~\bibnamefont {Lauterborn}},\ }\bibfield  {title}
  {\bibinfo {title} {{Membrane cleaning with ultrasonically driven bubbles}},\
  }\href {https://doi.org/10.1016/j.ultsonch.2016.12.012} {\bibfield  {journal}
  {\bibinfo  {journal} {Ultrasonics Sonochemistry}\ }\textbf {\bibinfo {volume}
  {37}},\ \bibinfo {pages} {542} (\bibinfo {year} {2017})}\BibitemShut
  {NoStop}%
\bibitem [{\citenamefont {{De La Torre}}(1992)}]{DeLaTorre1992}%
  \BibitemOpen
  \bibfield  {author} {\bibinfo {author} {\bibfnamefont {R.}~\bibnamefont {{De
  La Torre}}},\ }\emph {\bibinfo {title} {{Intravascular laser induced
  cavitation : a study of the mechanics with possible detrimental and
  beneficial effects}}},\ \href {http://dspace.mit.edu/handle/1721.1/17307}
  {Ph.D. thesis},\ \bibinfo  {school} {Massachusetts Institute of Technology}
  (\bibinfo {year} {1992})\BibitemShut {NoStop}%
\bibitem [{\citenamefont {Palanker}\ \emph {et~al.}(2002)\citenamefont
  {Palanker}, \citenamefont {Vankov},\ and\ \citenamefont
  {Miller}}]{Palanker2002}%
  \BibitemOpen
  \bibfield  {author} {\bibinfo {author} {\bibfnamefont {D.}~\bibnamefont
  {Palanker}}, \bibinfo {author} {\bibfnamefont {A.}~\bibnamefont {Vankov}},\
  and\ \bibinfo {author} {\bibfnamefont {J.}~\bibnamefont {Miller}},\
  }\bibfield  {title} {\bibinfo {title} {{Effect of the probe geometry on
  dynamics of cavitation}},\ }\href {https://doi.org/10.1117/12.472514}
  {\bibfield  {journal} {\bibinfo  {journal} {Proceedings of SPIE - The
  International Society for Optical Engineering}\ }\textbf {\bibinfo {volume}
  {4617}},\ \bibinfo {pages} {112} (\bibinfo {year} {2002})}\BibitemShut
  {NoStop}%
\bibitem [{\citenamefont {Brennen}(2003)}]{Brennen2003}%
  \BibitemOpen
  \bibfield  {author} {\bibinfo {author} {\bibfnamefont {C.~E.}\ \bibnamefont
  {Brennen}},\ }\bibfield  {title} {\bibinfo {title} {{Cavitation in Biological
  and Bioengineering Contexts}},\ }\href
  {https://authors.library.caltech.edu/73/} {\bibfield  {journal} {\bibinfo
  {journal} {Fifth International Symposium on Cavitation}\ } (\bibinfo {year}
  {2003})}\BibitemShut {NoStop}%
\bibitem [{\citenamefont {Canchi}\ \emph {et~al.}(2017)\citenamefont {Canchi},
  \citenamefont {Kelly}, \citenamefont {Hong}, \citenamefont {King},
  \citenamefont {Subhash},\ and\ \citenamefont {Sarntinoranont}}]{Canchi2017}%
  \BibitemOpen
  \bibfield  {author} {\bibinfo {author} {\bibfnamefont {S.}~\bibnamefont
  {Canchi}}, \bibinfo {author} {\bibfnamefont {K.}~\bibnamefont {Kelly}},
  \bibinfo {author} {\bibfnamefont {Y.}~\bibnamefont {Hong}}, \bibinfo {author}
  {\bibfnamefont {M.~A.}\ \bibnamefont {King}}, \bibinfo {author}
  {\bibfnamefont {G.}~\bibnamefont {Subhash}},\ and\ \bibinfo {author}
  {\bibfnamefont {M.}~\bibnamefont {Sarntinoranont}},\ }\bibfield  {title}
  {\bibinfo {title} {{Controlled single bubble cavitation collapse results in
  jet-induced injury in brain tissue}},\ }\href
  {https://doi.org/10.1016/j.jmbbm.2017.06.018} {\bibfield  {journal} {\bibinfo
   {journal} {Journal of the Mechanical Behavior of Biomedical Materials}\
  }\textbf {\bibinfo {volume} {74}},\ \bibinfo {pages} {261} (\bibinfo {year}
  {2017})}\BibitemShut {NoStop}%
\bibitem [{\citenamefont {{Oyarte G{\'{a}}lvez}}\ \emph
  {et~al.}(2020)\citenamefont {{Oyarte G{\'{a}}lvez}}, \citenamefont {Fraters},
  \citenamefont {Offerhaus}, \citenamefont {Versluis}, \citenamefont {Hunter},\
  and\ \citenamefont {{Fern{\'{a}}ndez Rivas}}}]{OyarteGalvez2020}%
  \BibitemOpen
  \bibfield  {author} {\bibinfo {author} {\bibfnamefont {L.}~\bibnamefont
  {{Oyarte G{\'{a}}lvez}}}, \bibinfo {author} {\bibfnamefont {A.}~\bibnamefont
  {Fraters}}, \bibinfo {author} {\bibfnamefont {H.~L.}\ \bibnamefont
  {Offerhaus}}, \bibinfo {author} {\bibfnamefont {M.}~\bibnamefont {Versluis}},
  \bibinfo {author} {\bibfnamefont {I.~W.}\ \bibnamefont {Hunter}},\ and\
  \bibinfo {author} {\bibfnamefont {D.}~\bibnamefont {{Fern{\'{a}}ndez
  Rivas}}},\ }\bibfield  {title} {\bibinfo {title} {{Microfluidics control the
  ballistic energy of thermocavitation liquid jets for needle-free
  injections}},\ }\href {https://doi.org/10.1063/1.5140264} {\bibfield
  {journal} {\bibinfo  {journal} {Journal of Applied Physics}\ }\textbf
  {\bibinfo {volume} {127}},\ \bibinfo {pages} {104901} (\bibinfo {year}
  {2020})}\BibitemShut {NoStop}%
\bibitem [{\citenamefont {Blake}\ \emph {et~al.}(1986)\citenamefont {Blake},
  \citenamefont {Taib},\ and\ \citenamefont {Doherty}}]{Blake1986}%
  \BibitemOpen
  \bibfield  {author} {\bibinfo {author} {\bibfnamefont {J.~R.}\ \bibnamefont
  {Blake}}, \bibinfo {author} {\bibfnamefont {B.~B.}\ \bibnamefont {Taib}},\
  and\ \bibinfo {author} {\bibfnamefont {G.}~\bibnamefont {Doherty}},\
  }\bibfield  {title} {\bibinfo {title} {{Transient cavities near boundaries.
  Part 1. Rigid boundary}},\ }\href {https://doi.org/10.1017/S0022112086000988}
  {\bibfield  {journal} {\bibinfo  {journal} {Journal of Fluid Mechanics}\
  }\textbf {\bibinfo {volume} {170}},\ \bibinfo {pages} {479} (\bibinfo {year}
  {1986})}\BibitemShut {NoStop}%
\bibitem [{\citenamefont {Brujan}\ \emph {et~al.}(2019)\citenamefont {Brujan},
  \citenamefont {Takahira},\ and\ \citenamefont {Ogasawara}}]{Brujan2019}%
  \BibitemOpen
  \bibfield  {author} {\bibinfo {author} {\bibfnamefont {E.~A.}\ \bibnamefont
  {Brujan}}, \bibinfo {author} {\bibfnamefont {H.}~\bibnamefont {Takahira}},\
  and\ \bibinfo {author} {\bibfnamefont {T.}~\bibnamefont {Ogasawara}},\
  }\bibfield  {title} {\bibinfo {title} {{Planar jets in collapsing cavitation
  bubbles}},\ }\href {https://doi.org/10.1016/j.expthermflusci.2018.10.007}
  {\bibfield  {journal} {\bibinfo  {journal} {Experimental Thermal and Fluid
  Science}\ }\textbf {\bibinfo {volume} {101}},\ \bibinfo {pages} {48}
  (\bibinfo {year} {2019})}\BibitemShut {NoStop}%
\bibitem [{\citenamefont {Zhang}\ \emph {et~al.}(2020)\citenamefont {Zhang},
  \citenamefont {Qiu}, \citenamefont {Zhang},\ and\ \citenamefont
  {Tang}}]{Zhang2020}%
  \BibitemOpen
  \bibfield  {author} {\bibinfo {author} {\bibfnamefont {Y.}~\bibnamefont
  {Zhang}}, \bibinfo {author} {\bibfnamefont {X.}~\bibnamefont {Qiu}}, \bibinfo
  {author} {\bibfnamefont {X.}~\bibnamefont {Zhang}},\ and\ \bibinfo {author}
  {\bibfnamefont {N.}~\bibnamefont {Tang}},\ }\bibfield  {title} {\bibinfo
  {title} {{Collapsing dynamics of a laser-induced cavitation bubble near the
  edge of a rigid wall}},\ }\href
  {https://doi.org/10.1016/J.ULTSONCH.2020.105157} {\bibfield  {journal}
  {\bibinfo  {journal} {Ultrasonics Sonochemistry}\ }\textbf {\bibinfo {volume}
  {67}},\ \bibinfo {pages} {105157} (\bibinfo {year} {2020})}\BibitemShut
  {NoStop}%
\bibitem [{\citenamefont {Benjamin}\ and\ \citenamefont
  {Ellis}(1966)}]{Benjamin1966}%
  \BibitemOpen
  \bibfield  {author} {\bibinfo {author} {\bibfnamefont {T.~B.}\ \bibnamefont
  {Benjamin}}\ and\ \bibinfo {author} {\bibfnamefont {A.~T.}\ \bibnamefont
  {Ellis}},\ }\bibfield  {title} {\bibinfo {title} {{The Collapse of Cavitation
  Bubbles and the Pressures thereby Produced against Solid Boundaries}},\
  }\href {http://www.jstor.org/stable/73553} {\bibfield  {journal} {\bibinfo
  {journal} {Philosophical Transactions of the Royal Society of London. Series
  A, Mathematical and Physical Sciences}\ }\textbf {\bibinfo {volume} {260}},\
  \bibinfo {pages} {221} (\bibinfo {year} {1966})}\BibitemShut {NoStop}%
\bibitem [{\citenamefont {Li}\ \emph {et~al.}(2015)\citenamefont {Li},
  \citenamefont {bo~Li},\ and\ \citenamefont {man Zhang}}]{Li2015}%
  \BibitemOpen
  \bibfield  {author} {\bibinfo {author} {\bibfnamefont {S.}~\bibnamefont
  {Li}}, \bibinfo {author} {\bibfnamefont {Y.~B.}\ \bibnamefont {Li}},\ and\
  \bibinfo {author} {\bibfnamefont {A.~M.}\ \bibnamefont {Zhang}},\ }\bibfield
  {title} {\bibinfo {title} {{Numerical analysis of the bubble jet impact on a
  rigid wall}},\ }\bibfield  {journal} {\bibinfo  {journal} {Applied Ocean
  Research}\ }\href {https://doi.org/10.1016/j.apor.2015.02.003}
  {10.1016/j.apor.2015.02.003} (\bibinfo {year} {2015})\BibitemShut {NoStop}%
\bibitem [{\citenamefont {Li}\ \emph {et~al.}(2016)\citenamefont {Li},
  \citenamefont {Han}, \citenamefont {Zhang},\ and\ \citenamefont
  {Wang}}]{Li2016}%
  \BibitemOpen
  \bibfield  {author} {\bibinfo {author} {\bibfnamefont {S.}~\bibnamefont
  {Li}}, \bibinfo {author} {\bibfnamefont {R.}~\bibnamefont {Han}}, \bibinfo
  {author} {\bibfnamefont {A.~M.}\ \bibnamefont {Zhang}},\ and\ \bibinfo
  {author} {\bibfnamefont {Q.~X.}\ \bibnamefont {Wang}},\ }\bibfield  {title}
  {\bibinfo {title} {{Analysis of pressure field generated by a collapsing
  bubble}},\ }\href {https://doi.org/10.1016/j.oceaneng.2016.03.016} {\bibfield
   {journal} {\bibinfo  {journal} {Ocean Engineering}\ }\textbf {\bibinfo
  {volume} {117}},\ \bibinfo {pages} {22} (\bibinfo {year} {2016})}\BibitemShut
  {NoStop}%
\bibitem [{\citenamefont {Supponen}\ \emph {et~al.}(2019)\citenamefont
  {Supponen}, \citenamefont {Obreschkow},\ and\ \citenamefont
  {Farhat}}]{Supponen2019}%
  \BibitemOpen
  \bibfield  {author} {\bibinfo {author} {\bibfnamefont {O.}~\bibnamefont
  {Supponen}}, \bibinfo {author} {\bibfnamefont {D.}~\bibnamefont
  {Obreschkow}},\ and\ \bibinfo {author} {\bibfnamefont {M.}~\bibnamefont
  {Farhat}},\ }\bibfield  {title} {\bibinfo {title} {{High-speed imaging of
  high pressures produced by cavitation bubbles}},\ }in\ \href
  {https://doi.org/10.1117/12.2523259} {\emph {\bibinfo {booktitle} {32nd
  International Congress on High-Speed Imaging and Photonics}}},\ Vol.\
  \bibinfo {volume} {11051},\ \bibinfo {editor} {edited by\ \bibinfo {editor}
  {\bibfnamefont {M.}~\bibnamefont {Versluis}}\ and\ \bibinfo {editor}
  {\bibfnamefont {E.}~\bibnamefont {Stride}}}\ (\bibinfo  {publisher} {SPIE},\
  \bibinfo {year} {2019})\ p.~\bibinfo {pages} {4}\BibitemShut {NoStop}%
\bibitem [{\citenamefont {Dijkink}\ and\ \citenamefont
  {Ohl}(2008)}]{Dijkink2008a}%
  \BibitemOpen
  \bibfield  {author} {\bibinfo {author} {\bibfnamefont {R.}~\bibnamefont
  {Dijkink}}\ and\ \bibinfo {author} {\bibfnamefont {C.~D.}\ \bibnamefont
  {Ohl}},\ }\bibfield  {title} {\bibinfo {title} {{Measurement of cavitation
  induced wall shear stress}},\ }\bibfield  {journal} {\bibinfo  {journal}
  {Applied Physics Letters}\ }\textbf {\bibinfo {volume} {93}},\ \href
  {https://doi.org/10.1063/1.3046735} {10.1063/1.3046735} (\bibinfo {year}
  {2008})\BibitemShut {NoStop}%
\bibitem [{\citenamefont {Koukouvinis}\ \emph {et~al.}(2018)\citenamefont
  {Koukouvinis}, \citenamefont {Strotos}, \citenamefont {Zeng}, \citenamefont
  {Gonzalez-Avila}, \citenamefont {Theodorakakos}, \citenamefont {Gavaises},\
  and\ \citenamefont {Ohl}}]{Koukouvinis2018}%
  \BibitemOpen
  \bibfield  {author} {\bibinfo {author} {\bibfnamefont {P.}~\bibnamefont
  {Koukouvinis}}, \bibinfo {author} {\bibfnamefont {G.}~\bibnamefont
  {Strotos}}, \bibinfo {author} {\bibfnamefont {Q.}~\bibnamefont {Zeng}},
  \bibinfo {author} {\bibfnamefont {S.~R.}\ \bibnamefont {Gonzalez-Avila}},
  \bibinfo {author} {\bibfnamefont {A.}~\bibnamefont {Theodorakakos}}, \bibinfo
  {author} {\bibfnamefont {M.}~\bibnamefont {Gavaises}},\ and\ \bibinfo
  {author} {\bibfnamefont {C.~D.}\ \bibnamefont {Ohl}},\ }\bibfield  {title}
  {\bibinfo {title} {{Parametric Investigations of the Induced Shear Stress by
  a Laser-Generated Bubble}},\ }\href
  {https://doi.org/10.1021/acs.langmuir.8b01274} {\bibfield  {journal}
  {\bibinfo  {journal} {Langmuir}\ }\textbf {\bibinfo {volume} {34}},\ \bibinfo
  {pages} {6428} (\bibinfo {year} {2018})}\BibitemShut {NoStop}%
\bibitem [{\citenamefont {Zeng}\ \emph {et~al.}(2018)\citenamefont {Zeng},
  \citenamefont {Gonzalez-Avila}, \citenamefont {Dijkink}, \citenamefont
  {Koukouvinis}, \citenamefont {Gavaises},\ and\ \citenamefont
  {Ohl}}]{Zeng2018}%
  \BibitemOpen
  \bibfield  {author} {\bibinfo {author} {\bibfnamefont {Q.}~\bibnamefont
  {Zeng}}, \bibinfo {author} {\bibfnamefont {S.~R.}\ \bibnamefont
  {Gonzalez-Avila}}, \bibinfo {author} {\bibfnamefont {R.}~\bibnamefont
  {Dijkink}}, \bibinfo {author} {\bibfnamefont {P.}~\bibnamefont
  {Koukouvinis}}, \bibinfo {author} {\bibfnamefont {M.}~\bibnamefont
  {Gavaises}},\ and\ \bibinfo {author} {\bibfnamefont {C.-D.}\ \bibnamefont
  {Ohl}},\ }\bibfield  {title} {\bibinfo {title} {{Wall shear stress from
  jetting cavitation bubbles}},\ }\href {https://doi.org/10.1017/jfm.2018.286}
  {\bibfield  {journal} {\bibinfo  {journal} {Journal of Fluid Mechanics}\
  }\textbf {\bibinfo {volume} {846}},\ \bibinfo {pages} {341} (\bibinfo {year}
  {2018})}\BibitemShut {NoStop}%
\bibitem [{\citenamefont {Gonzalez-Avila}\ \emph {et~al.}(2020)\citenamefont
  {Gonzalez-Avila}, \citenamefont {van Blokland}, \citenamefont {Zeng},\ and\
  \citenamefont {Ohl}}]{Gonzalez-Avila2020}%
  \BibitemOpen
  \bibfield  {author} {\bibinfo {author} {\bibfnamefont {S.~R.}\ \bibnamefont
  {Gonzalez-Avila}}, \bibinfo {author} {\bibfnamefont {A.~C.}\ \bibnamefont
  {van Blokland}}, \bibinfo {author} {\bibfnamefont {Q.}~\bibnamefont {Zeng}},\
  and\ \bibinfo {author} {\bibfnamefont {C.-D.}\ \bibnamefont {Ohl}},\
  }\bibfield  {title} {\bibinfo {title} {{Jetting and shear stress enhancement
  from cavitation bubbles collapsing in a narrow gap}},\ }\href
  {https://doi.org/10.1017/jfm.2019.938} {\bibfield  {journal} {\bibinfo
  {journal} {Journal of Fluid Mechanics}\ }\textbf {\bibinfo {volume} {884}},\
  \bibinfo {pages} {A23} (\bibinfo {year} {2020})}\BibitemShut {NoStop}%
\bibitem [{\citenamefont {Zeng}\ \emph {et~al.}(2022)\citenamefont {Zeng},
  \citenamefont {An},\ and\ \citenamefont {Ohl}}]{Zeng2022}%
  \BibitemOpen
  \bibfield  {author} {\bibinfo {author} {\bibfnamefont {Q.}~\bibnamefont
  {Zeng}}, \bibinfo {author} {\bibfnamefont {H.}~\bibnamefont {An}},\ and\
  \bibinfo {author} {\bibfnamefont {C.~D.}\ \bibnamefont {Ohl}},\ }\bibfield
  {title} {\bibinfo {title} {{Wall shear stress from jetting cavitation
  bubbles: influence of the stand-off distance and liquid viscosity}},\ }\href
  {https://doi.org/10.1017/JFM.2021.997} {\bibfield  {journal} {\bibinfo
  {journal} {Journal of Fluid Mechanics}\ }\textbf {\bibinfo {volume} {932}},\
  \bibinfo {pages} {14} (\bibinfo {year} {2022})}\BibitemShut {NoStop}%
\bibitem [{\citenamefont {Tagawa}\ \emph {et~al.}(2016)\citenamefont {Tagawa},
  \citenamefont {Yamamoto}, \citenamefont {Hayasaka},\ and\ \citenamefont
  {Kameda}}]{Tagawa2016}%
  \BibitemOpen
  \bibfield  {author} {\bibinfo {author} {\bibfnamefont {Y.}~\bibnamefont
  {Tagawa}}, \bibinfo {author} {\bibfnamefont {S.}~\bibnamefont {Yamamoto}},
  \bibinfo {author} {\bibfnamefont {K.}~\bibnamefont {Hayasaka}},\ and\
  \bibinfo {author} {\bibfnamefont {M.}~\bibnamefont {Kameda}},\ }\bibfield
  {title} {\bibinfo {title} {{On pressure impulse of a laser-induced underwater
  shock wave}},\ }\href {https://doi.org/10.1017/jfm.2016.644} {\bibfield
  {journal} {\bibinfo  {journal} {Journal of Fluid Mechanics}\ }\textbf
  {\bibinfo {volume} {808}},\ \bibinfo {pages} {5} (\bibinfo {year} {2016})} \BibitemShut
  {NoStop}%
\bibitem [{\citenamefont {Sinibaldi}\ \emph {et~al.}(2019)\citenamefont
  {Sinibaldi}, \citenamefont {Occhicone}, \citenamefont {{Alves Pereira}},
  \citenamefont {Caprini}, \citenamefont {Marino}, \citenamefont {Michelotti},\
  and\ \citenamefont {Casciola}}]{Sinibaldi2019}%
  \BibitemOpen
  \bibfield  {author} {\bibinfo {author} {\bibfnamefont {G.}~\bibnamefont
  {Sinibaldi}}, \bibinfo {author} {\bibfnamefont {A.}~\bibnamefont
  {Occhicone}}, \bibinfo {author} {\bibfnamefont {F.}~\bibnamefont {{Alves
  Pereira}}}, \bibinfo {author} {\bibfnamefont {D.}~\bibnamefont {Caprini}},
  \bibinfo {author} {\bibfnamefont {L.}~\bibnamefont {Marino}}, \bibinfo
  {author} {\bibfnamefont {F.}~\bibnamefont {Michelotti}},\ and\ \bibinfo
  {author} {\bibfnamefont {C.~M.}\ \bibnamefont {Casciola}},\ }\bibfield
  {title} {\bibinfo {title} {{Laser induced cavitation: Plasma generation and
  breakdown shockwave}},\ }\href {https://doi.org/10.1063/1.5119794} {\bibfield
   {journal} {\bibinfo  {journal} {Physics of Fluids}\ }\textbf {\bibinfo
  {volume} {31}},\ \bibinfo {pages} {103302} (\bibinfo {year}
  {2019})}\BibitemShut {NoStop}%
\bibitem [{\citenamefont {Trummler}\ \emph {et~al.}(2021)\citenamefont
  {Trummler}, \citenamefont {Schmidt},\ and\ \citenamefont
  {Adams}}]{Trummler2021}%
  \BibitemOpen
  \bibfield  {author} {\bibinfo {author} {\bibfnamefont {T.}~\bibnamefont
  {Trummler}}, \bibinfo {author} {\bibfnamefont {S.~J.}\ \bibnamefont
  {Schmidt}},\ and\ \bibinfo {author} {\bibfnamefont {N.~A.}\ \bibnamefont
  {Adams}},\ }\bibfield  {title} {\bibinfo {title} {{Numerical investigation of
  non-condensable gas effect on vapor bubble collapse}},\ }\href
  {https://doi.org/10.1063/5.0062399} {\bibfield  {journal} {\bibinfo
  {journal} {Physics of Fluids}\ }\textbf {\bibinfo {volume} {33}},\ \bibinfo
  {pages} {096107} (\bibinfo {year} {2021})} \BibitemShut {NoStop}%
\bibitem [{\citenamefont {Tagawa}\ and\ \citenamefont
  {Peters}(2018)}]{PhysRevFluids.3.081601}%
  \BibitemOpen
  \bibfield  {author} {\bibinfo {author} {\bibfnamefont {Y.}~\bibnamefont
  {Tagawa}}\ and\ \bibinfo {author} {\bibfnamefont {I.~R.}\ \bibnamefont
  {Peters}},\ }\bibfield  {title} {\bibinfo {title} {{Bubble collapse and jet
  formation in corner geometries}},\ }\href
  {https://doi.org/10.1103/PhysRevFluids.3.081601} {\bibfield  {journal}
  {\bibinfo  {journal} {Physical Review Fluids}\ }\textbf {\bibinfo {volume}
  {3}},\ \bibinfo {pages} {081601(R)} (\bibinfo {year} {2018})} \BibitemShut {NoStop}%
\bibitem [{\citenamefont {Molefe}\ and\ \citenamefont
  {Peters}(2019)}]{Molefe2019}%
  \BibitemOpen
  \bibfield  {author} {\bibinfo {author} {\bibfnamefont {L.}~\bibnamefont
  {Molefe}}\ and\ \bibinfo {author} {\bibfnamefont {I.~R.}\ \bibnamefont
  {Peters}},\ }\bibfield  {title} {\bibinfo {title} {{Jet direction in bubble
  collapse within rectangular and triangular channels}},\ }\href
  {https://doi.org/10.1103/PhysRevE.100.063105} {\bibfield  {journal} {\bibinfo
   {journal} {Physical Review E}\ }\textbf {\bibinfo {volume} {100}},\ \bibinfo
  {pages} {063105} (\bibinfo {year} {2019})}\BibitemShut {NoStop}%
\bibitem [{\citenamefont {Andrews}\ \emph
  {et~al.}(2020{\natexlab{a}})\citenamefont {Andrews}, \citenamefont {Rivas},\
  and\ \citenamefont {Peters}}]{Andrews2020}%
  \BibitemOpen
  \bibfield  {author} {\bibinfo {author} {\bibfnamefont {E.~D.}\ \bibnamefont
  {Andrews}}, \bibinfo {author} {\bibfnamefont {D.~F.}\ \bibnamefont {Rivas}},\
  and\ \bibinfo {author} {\bibfnamefont {I.~R.}\ \bibnamefont {Peters}},\
  }\bibfield  {title} {\bibinfo {title} {{Cavity collapse near slot
  geometries}},\ }\href {https://doi.org/10.1017/jfm.2020.552} {\bibfield
  {journal} {\bibinfo  {journal} {Journal of Fluid Mechanics}\ }\textbf
  {\bibinfo {volume} {901}},\ \bibinfo {pages} {29} (\bibinfo {year}
  {2020}{\natexlab{a}})} \BibitemShut {NoStop}%
\bibitem [{\citenamefont {Tomita}\ \emph {et~al.}(2002)\citenamefont {Tomita},
  \citenamefont {Robinson}, \citenamefont {Tong},\ and\ \citenamefont
  {Blake}}]{TOMITA2002}%
  \BibitemOpen
  \bibfield  {author} {\bibinfo {author} {\bibfnamefont {Y.}~\bibnamefont
  {Tomita}}, \bibinfo {author} {\bibfnamefont {P.~B.}\ \bibnamefont
  {Robinson}}, \bibinfo {author} {\bibfnamefont {R.~P.}\ \bibnamefont {Tong}},\
  and\ \bibinfo {author} {\bibfnamefont {J.~R.}\ \bibnamefont {Blake}},\
  }\bibfield  {title} {\bibinfo {title} {{Growth and collapse of cavitation
  bubbles near a curved rigid boundary}},\ }\href
  {https://doi.org/10.1017/S0022112002001209} {\bibfield  {journal} {\bibinfo
  {journal} {Journal of Fluid Mechanics}\ }\textbf {\bibinfo {volume} {466}},\
  \bibinfo {pages} {259} (\bibinfo {year} {2002})}\BibitemShut {NoStop}%
\bibitem [{\citenamefont {Zhang}\ \emph {et~al.}(2017)\citenamefont {Zhang},
  \citenamefont {Zhang}, \citenamefont {Wang},\ and\ \citenamefont
  {Cui}}]{Zhang2017}%
  \BibitemOpen
  \bibfield  {author} {\bibinfo {author} {\bibfnamefont {S.}~\bibnamefont
  {Zhang}}, \bibinfo {author} {\bibfnamefont {A.~M.}\ \bibnamefont {Zhang}},
  \bibinfo {author} {\bibfnamefont {S.~P.}\ \bibnamefont {Wang}},\ and\
  \bibinfo {author} {\bibfnamefont {J.}~\bibnamefont {Cui}},\ }\bibfield
  {title} {\bibinfo {title} {{Dynamic characteristics of large scale spark
  bubbles close to different boundaries}},\ }\href
  {https://doi.org/10.1063/1.4986821} {\bibfield  {journal} {\bibinfo
  {journal} {Physics of Fluids}\ }\textbf {\bibinfo {volume} {29}},\ \bibinfo
  {pages} {092107} (\bibinfo {year} {2017})}\BibitemShut {NoStop}%
\bibitem [{\citenamefont {Quah}\ \emph {et~al.}(2018)\citenamefont {Quah},
  \citenamefont {Karri}, \citenamefont {Ohl}, \citenamefont {Klaseboer},\ and\
  \citenamefont {Khoo}}]{Quah2018}%
  \BibitemOpen
  \bibfield  {author} {\bibinfo {author} {\bibfnamefont {E.~W.}\ \bibnamefont
  {Quah}}, \bibinfo {author} {\bibfnamefont {B.}~\bibnamefont {Karri}},
  \bibinfo {author} {\bibfnamefont {S.-W.}\ \bibnamefont {Ohl}}, \bibinfo
  {author} {\bibfnamefont {E.}~\bibnamefont {Klaseboer}},\ and\ \bibinfo
  {author} {\bibfnamefont {B.~C.}\ \bibnamefont {Khoo}},\ }\bibfield  {title}
  {\bibinfo {title} {{Expansion and collapse of an initially off-centered
  bubble within a narrow gap and the effect of a free surface}},\ }\href
  {https://doi.org/10.1016/J.IJMULTIPHASEFLOW.2017.09.013} {\bibfield
  {journal} {\bibinfo  {journal} {International Journal of Multiphase Flow}\
  }\textbf {\bibinfo {volume} {99}},\ \bibinfo {pages} {62} (\bibinfo {year}
  {2018})}\BibitemShut {NoStop}%
\bibitem [{\citenamefont {Kiyama}\ \emph {et~al.}(2021)\citenamefont {Kiyama},
  \citenamefont {Shimazaki}, \citenamefont {Gordillo},\ and\ \citenamefont
  {Tagawa}}]{Kiyama2021}%
  \BibitemOpen
  \bibfield  {author} {\bibinfo {author} {\bibfnamefont {A.}~\bibnamefont
  {Kiyama}}, \bibinfo {author} {\bibfnamefont {T.}~\bibnamefont {Shimazaki}},
  \bibinfo {author} {\bibfnamefont {J.~M.}\ \bibnamefont {Gordillo}},\ and\
  \bibinfo {author} {\bibfnamefont {Y.}~\bibnamefont {Tagawa}},\ }\bibfield
  {title} {\bibinfo {title} {{The direction of the microjet produced by the
  collapse of a cavitation bubble locatedin between a wall and a free
  surface}},\ }\href {https://doi.org/10.1103/PhysRevFluids.6.083601}
  {\bibfield  {journal} {\bibinfo  {journal} {Physical Review Fluids}\ }\textbf
  {\bibinfo {volume} {6}},\ \bibinfo {pages} {083601} (\bibinfo {year}
  {2021})}
  \BibitemShut {NoStop}%
\bibitem [{\citenamefont {Han}\ \emph {et~al.}(2018)\citenamefont {Han},
  \citenamefont {Zhu}, \citenamefont {Guo}, \citenamefont {Liu},\ and\
  \citenamefont {Ni}}]{Han2018}%
  \BibitemOpen
  \bibfield  {author} {\bibinfo {author} {\bibfnamefont {B.}~\bibnamefont
  {Han}}, \bibinfo {author} {\bibfnamefont {R.}~\bibnamefont {Zhu}}, \bibinfo
  {author} {\bibfnamefont {Z.}~\bibnamefont {Guo}}, \bibinfo {author}
  {\bibfnamefont {L.}~\bibnamefont {Liu}},\ and\ \bibinfo {author}
  {\bibfnamefont {X.~W.}\ \bibnamefont {Ni}},\ }\bibfield  {title} {\bibinfo
  {title} {{Control of the liquid jet formation through the symmetric and
  asymmetric collapse of a single bubble generated between two parallel solid
  plates}},\ }\href {https://doi.org/10.1016/j.euromechflu.2018.05.003}
  {\bibfield  {journal} {\bibinfo  {journal} {European Journal of Mechanics,
  B/Fluids}\ }\textbf {\bibinfo {volume} {72}},\ \bibinfo {pages} {114}
  (\bibinfo {year} {2018})}\BibitemShut {NoStop}%
\bibitem [{\citenamefont {Kr{\"{o}}ninger}\ \emph {et~al.}(2010)\citenamefont
  {Kr{\"{o}}ninger}, \citenamefont {K{\"{o}}hler}, \citenamefont {Kurz},\ and\
  \citenamefont {Lauterborn}}]{Kroninger2010}%
  \BibitemOpen
  \bibfield  {author} {\bibinfo {author} {\bibfnamefont {D.}~\bibnamefont
  {Kr{\"{o}}ninger}}, \bibinfo {author} {\bibfnamefont {K.}~\bibnamefont
  {K{\"{o}}hler}}, \bibinfo {author} {\bibfnamefont {T.}~\bibnamefont {Kurz}},\
  and\ \bibinfo {author} {\bibfnamefont {W.}~\bibnamefont {Lauterborn}},\
  }\bibfield  {title} {\bibinfo {title} {{Particle tracking velocimetry of the
  flow field around a collapsing cavitation bubble}},\ }\href
  {https://doi.org/10.1007/s00348-009-0743-1} {\bibfield  {journal} {\bibinfo
  {journal} {Experiments in Fluids}\ }\textbf {\bibinfo {volume} {48}},\
  \bibinfo {pages} {395} (\bibinfo {year} {2010})}\BibitemShut {NoStop}%
\bibitem [{\citenamefont {Supponen}\ \emph {et~al.}(2016)\citenamefont
  {Supponen}, \citenamefont {Obreschkow}, \citenamefont {Tinguely},
  \citenamefont {Kobel}, \citenamefont {Dorsaz},\ and\ \citenamefont
  {Farhat}}]{Supponen2016}%
  \BibitemOpen
  \bibfield  {author} {\bibinfo {author} {\bibfnamefont {O.}~\bibnamefont
  {Supponen}}, \bibinfo {author} {\bibfnamefont {D.}~\bibnamefont
  {Obreschkow}}, \bibinfo {author} {\bibfnamefont {M.}~\bibnamefont
  {Tinguely}}, \bibinfo {author} {\bibfnamefont {P.}~\bibnamefont {Kobel}},
  \bibinfo {author} {\bibfnamefont {N.}~\bibnamefont {Dorsaz}},\ and\ \bibinfo
  {author} {\bibfnamefont {M.}~\bibnamefont {Farhat}},\ }\bibfield  {title}
  {\bibinfo {title} {{Scaling laws for jets of single cavitation bubbles}},\
  }\href {https://doi.org/10.1017/jfm.2016.463} {\bibfield  {journal} {\bibinfo
   {journal} {Journal of Fluid Mechanics}\ }\textbf {\bibinfo {volume} {802}},\
  \bibinfo {pages} {263} (\bibinfo {year} {2016})}\BibitemShut {NoStop}%
\bibitem [{\citenamefont {Blake}\ and\ \citenamefont
  {Cerone}(1982)}]{Blake1982}%
  \BibitemOpen
  \bibfield  {author} {\bibinfo {author} {\bibfnamefont {J.~R.}\ \bibnamefont
  {Blake}}\ and\ \bibinfo {author} {\bibfnamefont {P.}~\bibnamefont {Cerone}},\
  }\bibfield  {title} {\bibinfo {title} {{A note on the impulse due to a vapour
  bubble near a boundary}},\ }\href {https://doi.org/10.1017/s0334270000000321}
  {\bibfield  {journal} {\bibinfo  {journal} {The Journal of the Australian
  Mathematical Society. Series B. Applied Mathematics}\ }\textbf {\bibinfo
  {volume} {23}},\ \bibinfo {pages} {383} (\bibinfo {year} {1982})}\BibitemShut
  {NoStop}%
\bibitem [{\citenamefont {Kucera}\ and\ \citenamefont
  {Blake}(1990)}]{Kucera1990}%
  \BibitemOpen
  \bibfield  {author} {\bibinfo {author} {\bibfnamefont {A.}~\bibnamefont
  {Kucera}}\ and\ \bibinfo {author} {\bibfnamefont {J.~R.}\ \bibnamefont
  {Blake}},\ }\bibfield  {title} {\bibinfo {title} {{Approximate methods for
  modelling cavitation bubbles near boundaries}},\ }\href
  {https://doi.org/10.1017/S0004972700017834} {\bibfield  {journal} {\bibinfo
  {journal} {Bulletin of the Australian Mathematical Society}\ }\textbf
  {\bibinfo {volume} {41}},\ \bibinfo {pages} {1} (\bibinfo {year}
  {1990})}\BibitemShut {NoStop}%
\bibitem [{\citenamefont {Harris}(1996)}]{Harris1996}%
  \BibitemOpen
  \bibfield  {author} {\bibinfo {author} {\bibfnamefont {P.~J.}\ \bibnamefont
  {Harris}},\ }\bibfield  {title} {\bibinfo {title} {{The numerical
  determination of the Kelvin impulse of a bubble close to a submerged rigid
  structure}},\ }\href {https://doi.org/10.1016/0045-7825(95)00914-0}
  {\bibfield  {journal} {\bibinfo  {journal} {Computer Methods in Applied
  Mechanics and Engineering}\ }\textbf {\bibinfo {volume} {130}},\ \bibinfo
  {pages} {195} (\bibinfo {year} {1996})}\BibitemShut {NoStop}%
\bibitem [{\citenamefont {Blake}\ \emph {et~al.}(2015)\citenamefont {Blake},
  \citenamefont {Leppinen},\ and\ \citenamefont {Wang}}]{Blake2015}%
  \BibitemOpen
  \bibfield  {author} {\bibinfo {author} {\bibfnamefont {J.~R.}\ \bibnamefont
  {Blake}}, \bibinfo {author} {\bibfnamefont {D.~M.}\ \bibnamefont
  {Leppinen}},\ and\ \bibinfo {author} {\bibfnamefont {Q.}~\bibnamefont
  {Wang}},\ }\bibfield  {title} {\bibinfo {title} {{Cavitation and bubble
  dynamics: The Kelvin impulse and its applications}},\ }\href
  {https://doi.org/10.1098/rsfs.2015.0017} {\bibfield  {journal} {\bibinfo
  {journal} {Interface Focus}\ }\textbf {\bibinfo {volume} {5}},\ \bibinfo
  {pages} {1} (\bibinfo {year} {2015})}\BibitemShut {NoStop}%
\bibitem [{\citenamefont {Supponen}\ \emph {et~al.}(2017)\citenamefont
  {Supponen}, \citenamefont {Obreschkow}, \citenamefont {Kobel}, \citenamefont
  {Tinguely}, \citenamefont {Dorsaz},\ and\ \citenamefont
  {Farhat}}]{Supponen2017}%
  \BibitemOpen
  \bibfield  {author} {\bibinfo {author} {\bibfnamefont {O.}~\bibnamefont
  {Supponen}}, \bibinfo {author} {\bibfnamefont {D.}~\bibnamefont
  {Obreschkow}}, \bibinfo {author} {\bibfnamefont {P.}~\bibnamefont {Kobel}},
  \bibinfo {author} {\bibfnamefont {M.}~\bibnamefont {Tinguely}}, \bibinfo
  {author} {\bibfnamefont {N.}~\bibnamefont {Dorsaz}},\ and\ \bibinfo {author}
  {\bibfnamefont {M.}~\bibnamefont {Farhat}},\ }\bibfield  {title} {\bibinfo
  {title} {{Shock waves from nonspherical cavitation bubbles}},\ }\href
  {https://doi.org/10.1103/PhysRevFluids.2.093601} {\bibfield  {journal}
  {\bibinfo  {journal} {Physical Review Fluids}\ }\textbf {\bibinfo {volume}
  {2}},\ \bibinfo {pages} {093601} (\bibinfo {year} {2017})} \BibitemShut {NoStop}%
\bibitem [{\citenamefont {Supponen}\ \emph {et~al.}(2018)\citenamefont
  {Supponen}, \citenamefont {Obreschkow},\ and\ \citenamefont
  {Farhat}}]{Supponen2018}%
  \BibitemOpen
  \bibfield  {author} {\bibinfo {author} {\bibfnamefont {O.}~\bibnamefont
  {Supponen}}, \bibinfo {author} {\bibfnamefont {D.}~\bibnamefont
  {Obreschkow}},\ and\ \bibinfo {author} {\bibfnamefont {M.}~\bibnamefont
  {Farhat}},\ }\bibfield  {title} {\bibinfo {title} {{Rebounds of deformed
  cavitation bubbles}},\ }\href
  {https://doi.org/10.1103/PhysRevFluids.3.103604} {\bibfield  {journal}
  {\bibinfo  {journal} {Physical Review Fluids}\ }\textbf {\bibinfo {volume}
  {3}},\ \bibinfo {pages} {103604} (\bibinfo {year} {2018})} \BibitemShut {NoStop}%
\bibitem [{\citenamefont {Blake}(1988)}]{Blake1988}%
  \BibitemOpen
  \bibfield  {author} {\bibinfo {author} {\bibfnamefont {J.~R.}\ \bibnamefont
  {Blake}},\ }\bibfield  {title} {\bibinfo {title} {{The Kelvin impulse:
  application to cavitation bubble dynamics}},\ }\href
  {https://doi.org/10.1017/s0334270000006111} {\bibfield  {journal} {\bibinfo
  {journal} {The Journal of the Australian Mathematical Society. Series B.
  Applied Mathematics}\ }\textbf {\bibinfo {volume} {30}},\ \bibinfo {pages}
  {127} (\bibinfo {year} {1988})}\BibitemShut {NoStop}%
\bibitem [{\citenamefont {Brennen}(1995)}]{Brennen1995}%
  \BibitemOpen
  \bibfield  {author} {\bibinfo {author} {\bibfnamefont {C.~E.}\ \bibnamefont
  {Brennen}},\ }\href {https://resolver.caltech.edu/CaltechBOOK:1995.001}
  {\emph {\bibinfo {title} {{Cavitation and Bubble Dynamics}}}}\ (\bibinfo
  {publisher} {Oxford University Press},\ \bibinfo {year} {1995})\BibitemShut
  {NoStop}%
\bibitem [{\citenamefont {Obreschkow}\ \emph {et~al.}(2012)\citenamefont
  {Obreschkow}, \citenamefont {Bruderer},\ and\ \citenamefont
  {Farhat}}]{Obreschkow2012}%
  \BibitemOpen
  \bibfield  {author} {\bibinfo {author} {\bibfnamefont {D.}~\bibnamefont
  {Obreschkow}}, \bibinfo {author} {\bibfnamefont {M.}~\bibnamefont
  {Bruderer}},\ and\ \bibinfo {author} {\bibfnamefont {M.}~\bibnamefont
  {Farhat}},\ }\bibfield  {title} {\bibinfo {title} {{Analytical approximations
  for the collapse of an empty spherical bubble}},\ }\href
  {https://doi.org/10.1103/PhysRevE.85.066303} {\bibfield  {journal} {\bibinfo
  {journal} {Physical Review E - Statistical, Nonlinear, and Soft Matter
  Physics}\ }\textbf {\bibinfo {volume} {85}},\ \bibinfo {pages} {066303}
  (\bibinfo {year} {2012})} \BibitemShut {NoStop}%
\bibitem [{\citenamefont {Peters}\ and\ \citenamefont
  {Tagawa}(2018)}]{cornersdata}%
  \BibitemOpen
  \bibfield  {author} {\bibinfo {author} {\bibfnamefont {I.}~\bibnamefont
  {Peters}}\ and\ \bibinfo {author} {\bibfnamefont {Y.}~\bibnamefont
  {Tagawa}},\ }\href {https://dx.doi.org/10.5258/SOTON/D0595} {\bibinfo {title}
  {{Dataset for Bubble Collapse and Jet Formation in Corner Geometries}}}
  (\bibinfo {year} {2018})\BibitemShut {NoStop}%
\bibitem [{\citenamefont {Peters}\ and\ \citenamefont
  {Molefe}(2019)}]{prismsdata}%
  \BibitemOpen
  \bibfield  {author} {\bibinfo {author} {\bibfnamefont {I.}~\bibnamefont
  {Peters}}\ and\ \bibinfo {author} {\bibfnamefont {L.}~\bibnamefont
  {Molefe}},\ }\href {https://dx.doi.org/10.5258/SOTON/D1151} {\bibinfo {title}
  {{Dataset for Jet direction in bubble collapse within rectangular and
  triangular channels}}} (\bibinfo {year} {2019})\BibitemShut {NoStop}%
\bibitem [{\citenamefont {Andrews}\ \emph
  {et~al.}(2020{\natexlab{b}})\citenamefont {Andrews}, \citenamefont {Rivas},\
  and\ \citenamefont {Peters}}]{slotsdata}%
  \BibitemOpen
  \bibfield  {author} {\bibinfo {author} {\bibfnamefont {E.~D.}\ \bibnamefont
  {Andrews}}, \bibinfo {author} {\bibfnamefont {D.~F.}\ \bibnamefont {Rivas}},\
  and\ \bibinfo {author} {\bibfnamefont {I.}~\bibnamefont {Peters}},\ }\href
  {http://dx.doi.org/10.5258/SOTON/D1454} {\bibinfo {title} {{Dataset for
  Cavity collapse near slot geometries}}} (\bibinfo {year}
  {2020}{\natexlab{b}})\BibitemShut {NoStop}%
\bibitem [{\citenamefont {Tinguely}\ \emph {et~al.}(2012)\citenamefont
  {Tinguely}, \citenamefont {Obreschkow}, \citenamefont {Kobel}, \citenamefont
  {Dorsaz}, \citenamefont {{de Bosset}},\ and\ \citenamefont
  {Farhat}}]{Tinguely2012}%
  \BibitemOpen
  \bibfield  {author} {\bibinfo {author} {\bibfnamefont {M.}~\bibnamefont
  {Tinguely}}, \bibinfo {author} {\bibfnamefont {D.}~\bibnamefont
  {Obreschkow}}, \bibinfo {author} {\bibfnamefont {P.}~\bibnamefont {Kobel}},
  \bibinfo {author} {\bibfnamefont {N.}~\bibnamefont {Dorsaz}}, \bibinfo
  {author} {\bibfnamefont {A.}~\bibnamefont {{de Bosset}}},\ and\ \bibinfo
  {author} {\bibfnamefont {M.}~\bibnamefont {Farhat}},\ }\bibfield  {title}
  {\bibinfo {title} {{Energy partition at the collapse of spherical cavitation
  bubbles}},\ }\href {https://doi.org/10.1103/PhysRevE.86.046315} {\bibfield
  {journal} {\bibinfo  {journal} {Physical Review E - Statistical, Nonlinear,
  and Soft Matter Physics}\ }\textbf {\bibinfo {volume} {86}},\ \bibinfo
  {pages} {046315} (\bibinfo {year} {2012})} \BibitemShut {NoStop}%
\end{thebibliography}

%

\end{document}